\documentclass[12pt,preprint]{aastex}
\usepackage{natbib}
\usepackage{graphicx}

\shorttitle{Spectroscopy of PNe in M31}
\shortauthors{X. Fang et al.}

\begin{document}

\title{Spectroscopic Observations of Planetary Nebulae in the Northern Spur
of M31}

\author{X. Fang$^1$,
        Y. Zhang$^2$,
        R. Garc\'{i}a-Benito$^3$,
        X.-W. Liu$^{1,4}$, and
        H.-B. Yuan$^{4,5}$\\
        $^1$Department of Astronomy, School of Physics, Peking University,
        Beijing 100871, China\\ %fangx@pku.edu.cn\\
        $^2$Department of Physics, University of Hong Kong, Hong Kong\\
        $^3$Instituto de Astrof\'{i}sica de Andaluc\'{i}a (CSIC), Glorieta
            de la Astronom\'{i}a s/n, E-18008 Granada, Spain\\
        $^4$Kavli Institute for Astronomy and Astrophysics, Peking
        University, Beijing 100871, China\\
        $^5$LAMOST Fellow}
\email{fangx@pku.edu.cn}

%\date{Accepted . Received }

%\pagerange{\pageref{firstpage}--\pageref{lastpage}} \pubyear{2013}

%\maketitle

\begin{abstract}
We present spectroscopy of three planetary nebulae (PNe) in the Northern Spur
of the Andromeda Galaxy (M31) obtained with the Double Spectrograph on the
5.1\,m Hale Telescope at the Palomar Observatory. The samples are
selected from the observations of Merrett et al. Our purpose is to
investigate formation of the substructures of M31 using PNe as a tracer of
chemical abundances. The [O~{\sc iii}] $\lambda$4363 line is detected in the
spectra of two objects, enabling temperature determinations. Ionic abundances
are derived from the observed collisionally excited lines, and elemental
abundances of nitrogen, oxygen, and neon as well as sulphur and argon are
estimated. Correlations between oxygen and $\alpha$-element abundance ratios
are studied, using our sample and the M31 disk and bulge PNe from the
literature. In one of the three PNe, we observed relatively higher oxygen
abundance compared to the disk sample in M31 at similar galactocentric
distances. The results of at least one of the three Northern Spur
PNe might be in line with the proposed possible origin of the Northern Spur
substructure of M31, i.e. the Northern Spur is connected to the Southern
Stream and both substructures comprise the tidal debris of the satellite
galaxies of M31.
%However, more observations of the PNe in different substructures are needed
%to sketch a complete picture of the formation and evolution processes of the
%galaxy.
\end{abstract}


\section{\label{part1}
Introduction}

In hierarchical cosmology, accretion (or merging) of smaller galaxies
contributes significantly to the growth of a large galaxy (e.g. the classical
theoretical work by \citealt{whi78} and \citealt{wr78}), especially in its
early evolutionary stage. The most recent observational studies of the growth
of galaxies were carried out by the CALIFA\footnote{Calar Alto Legacy Integral
Field spectroscopy Area Survey. URL: http://califa.caha.es} group
\citep{per13}, who analyzed 105 galaxies of the largest three-dimensional
spectroscopic survey of galaxies in the local universe. The outskirts of
galaxies hold fundamental clues about their formation history. It is into
these regions that new material continues to arrive as part of the ongoing
assembly and was deposited during the violent interactions in the galaxy's
distant past. Since it takes a very long time for the accreted material to be
erased by the process of phase mixing due to the long dynamical timescales,
we can detect various substructures, e.g. tidal tails, that are expected to
be harbored in the areas beyond the disk of a large galaxy. The tidal
disruption of these small systems is expected to result in loosely bounded
stars surrounding the galaxy, at distances up to 10\,--\,100 times the radius
of the central disk \citep{ans06}. Substructures like stellar streams, have
been observed in both Milky Way (MW) and other large spiral galaxies. The
Sagittarius Dwarf Galaxy provides a fine example in the MW (e.g.
\citealt{iba01a}; \citealt{maj03}).

The Andromeda Galaxy (M31) is the nearest large spiral system and one of the
best candidates for studying the debris of interaction. A large number of
coherent stellar substructures have been detected in its halo and outer disk
(e.g. \citealt{iba01b,iba07}; \citealt{fer02}; \citealt{irw05};
\citealt{mccon03,mccon04,mccon09}). The number, luminosity, morphology and
stellar population of the relics provide important clues to the assemblage
history of the galaxy. However, a comprehensive survey of those relics is
quite difficult, given their intrinsic faintness, dense distribution and the
vast space over which the stars are spread. Those factors make the stars and
clusters in M31 not ideal tools for abundance studies. Besides, there is also
contamination from the foreground stars of the MW. Planetary nebulae
(PNe) are excellent tracers to study the chemistry, kinematics, and stellar
contents of the substructures because they are bright but not packed too
closely. PNe are easily detectable at the distance of M31 (785~kpc;
\citealt{mccon05}), given their bright narrow emission line spectra.
They are also one of the best candidates that can provide both very
accurate velocities and precision abundance measurements of elements such as
O, He, Ne, N, Ar and S. Observations and chemical studies of PNe (as
well as H~{\sc ii} regions) in M31 have been carried out decades ago (e.g.
\citealt{fj78a, fj78b}; \citealt{jf86}; \citealt{srm98}; \citealt{rsm99};
\citealt{jc99}; the most recent observations are by \citealt{kwi12},
\citealt{zb12}, and \citealt{san12}). Using the nebular sample available,
oxygen abundance gradient in M31 has been derived. That helps to understand
the chemical evolution of the galaxy.

A number of substructures have been observed in the outer halo of M31.
The Northern Spur is a peculiar low surface brightness structure sticking
out of M31's disk, which contains a metal-rich stellar population. It had
been observed to lie in the direction of M31's gaseous warp decades ago (e.g.
\citealt{ne77}). \citet{inn82} noticed the anti-symmetrical warping of the
stellar disk along the major axis of M31 through digital stacking of Palomar
Schmidt plates of the galaxy. \citet{wk88} observed a faint light bending
away from the northern major axis using multi-color photometry and attributed
that to either a possible galactic reflection nebula or a warp in the outer
stellar disk of M31. However, the results for the northern outer disk of M31,
according to \citet{wk88}, was not conclusive due to faintness of the emission
there. The warp observed by both \citet{inn82} and \citet{wk88} to the
northeastern major axis of M31 starts at about 18~kpc from the galactic
center and extends to the region well beyond $\sim$25~kpc. Since the Northern
Spur lies in the direction of M31's gaseous warp, its projection away from
the galactic plane was usually attributed to a severe warp in the stellar disk.

Progress in observational and data-processing techniques during the past
decade have enabled large surveys. The faint substructures of M31 can now
be studied in greater details, and consequently their properties be better
understood. \citet{iba01b} discovered a giant stream (i.e. Southern Stream)
of metal-rich stars within the halo of M31, and they attributed the possible
source of the stream to the dwarf galaxies M32 and NGC\,205, which are close
companions of M31 and might have lost a substantial number of stars due to
tidal interactions. \citet{iba01b} concluded that the epoch of galaxy
building of M31 still continues, and that tidal streams might be a generic
feature of galaxy haloes.
\citet{fer02} carried out a panoramic survey of the halo and outer disk of
M31 and studied the density and color distribution of red giant branch (RGB)
stars. They confirmed the Southern Stream first announced by \citet{iba01b}
and found enhancement in both stellar density and metallicity in the Northern
Spur. Inspired by the asymmetry of the Southern Stream (i.e. it does not
appear to the north-west), \citet{fer02} hypothesized that the Southern
Stream might be associated with Northern Spur. \citet{mccon03} measured a
radial distance change along the Stream by analyzing the metal-rich RGB
luminosity function, and derived an angle of 60$^{\rm o}$ of the stream to
the line of sight. Combining the distance gradient with the angular extent,
\citet{mccon03} found that the Southern Stream extends from approximately
100~kpc behind to 40~kpc in front of M31. \citet{mer03} proposed that Southern
Stream might be connected to the Northern Spur and presented a
three-dimensional orbit for that connection by studying the dynamics of PNe
in the disk of M31. They also suggested that M32 might be the source of the
two substructures mainly because the satellite is found to coincide with the
Southern Stream in both the spatial position and velocity.

Currently, the exact origin of the Northern Spur is still largely
unknown, and the dynamical model proposed by \citet{mer03} needs to be tested.
However, the observational data in this region is scarce.
In order to assess the hypothesis of the origin of the Northern Spur by
studying its chemistry, we present spectroscopic observations of PNe in this
substructure and deduce their elemental abundances from emission
lines detected in the spectra. This is the first chemical study of the PNe
in the Northern Spur with spectroscopy. Section\,\ref{part2}
of this paper presents observations and data reduction, and demonstrates the
challenge we have in observing the PNe in the Northern Spur.  Line flux
measurements, plasma diagnostics and abundance determinations are given in
Section\,\ref{part3}. Comparison of the abundances in our sample with those
in the M31 disk and bulge PNe (also H~{\sc ii} regions) is presented in
Section\,\ref{part4}. Discussion of the possible origin of the Northern Spur
is also given in Section\,\ref{part4}. Summary and conclusion are given in
Section\,\ref{part5}.

\section{\label{part2}
Observations and data reduction}

\subsection{\label{part2:a}
Target selection}

\begin{figure*}
\begin{center}
\includegraphics[width=12cm,angle=0]{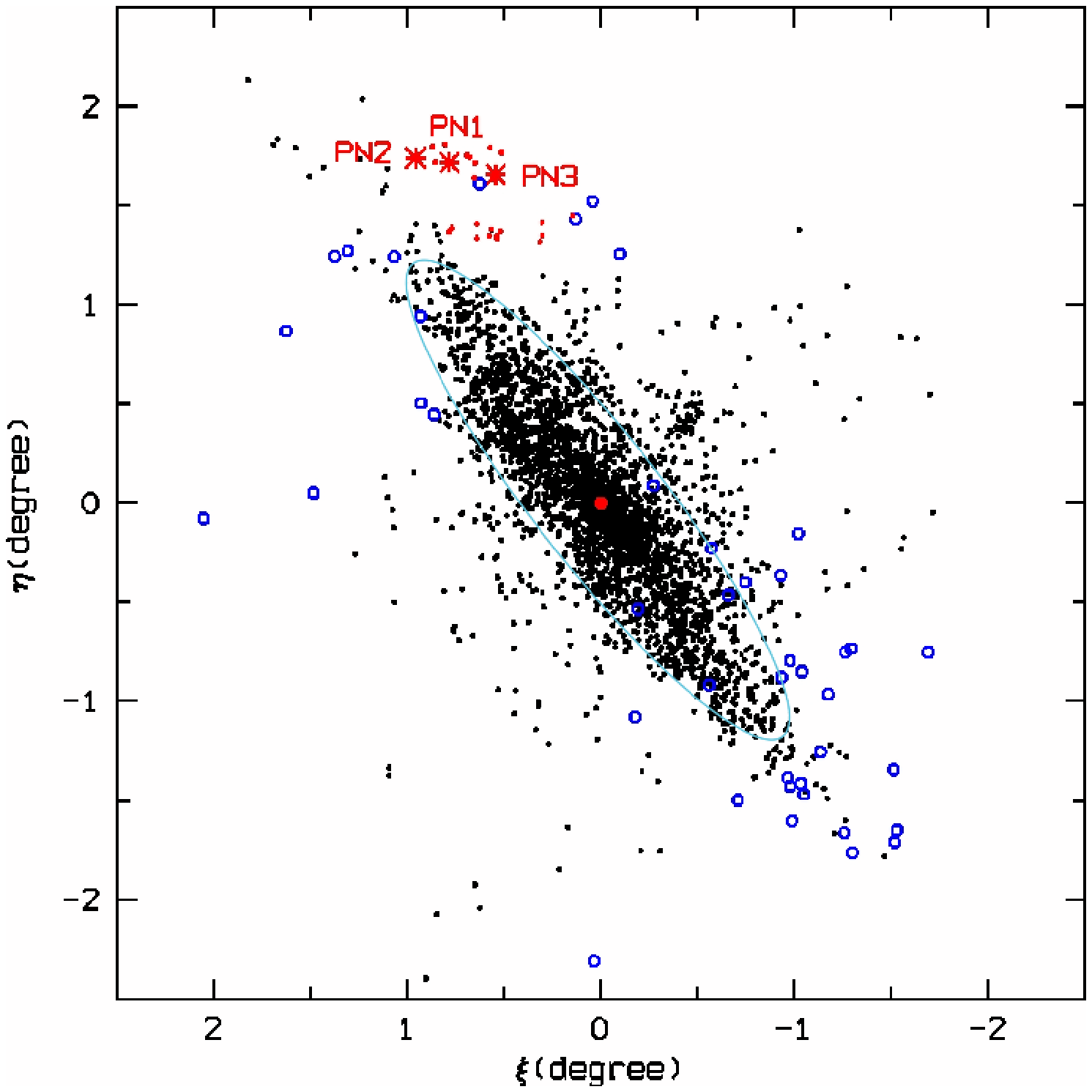}
\caption{M31 PNe observed by \citet[black dots]{mer06} and LAMOST \citep[blue
open circles]{yuan10}. The small red dots to the north of M31 disk are the
PNe in the Northern Spur identified by \citet{mer06}. The three Northern Spur
PNe, namely PN1, PN2 and PN3, observed and studied in the current paper are
indicated by red asterisks. The center of M31 is indicated with a
red dot. The coordinates $\xi$ and $\eta$ represent the M31-based reference
frame defined by \citet{hbk91}. The green ellipse is a disk with a 2$^{\rm
o}$ radius (27.4~kpc) around the center of M31, assuming an inclination angle
of 77.7$^{\rm o}$ and a position angle of 37.7$^{\rm o}$ \citep{dev58} for
the M31 disk.}
\label{m31_pne}
\end{center}
\end{figure*}

The main purpose of the current observation is to try to understand the
formation history of substructure in the outer disk of M31 by
studying the chemistry of PNe in the Northern Spur of M31. In order to obtain
accurate measurements of the heavy element abundances, we need to know
the physical conditions, most importantly, the electron temperatures
of those PNe. However, the heliocentric velocity of M31 ($\sim\,-$300\,km/s),
combined with the very bright mercury line Hg~{\sc i}
$\lambda$4358.34\footnote{Wavelength of the mercury line is adopted from the
National Institute of Standards and Technology (NIST) Atomic Spectra Database.
URL: http://www.nist.gov/pml/data/asd.cfm. The data source is \citet{ssr96}.}
from the nearby urban area, makes the [O~{\sc iii}] $\lambda$4363.21 auroral
line difficult to resolve. In order to make the [O~{\sc iii}] $\lambda$4363
line measurable, PNe with lowest radial velocities are favored. Candidates for
observations were selected from \citet{mer06}. Most of the candidates have
either low radial velocities but also low surface brightness (e.g. with the
magnitudes of [O~{\sc iii}] $\lambda$5007 $m_{\rm 5007}>$21.5), or relatively
high surface brightness but also high radial velocities. Finally, a compromise
between magnitude of the [O~{\sc iii}] $\lambda$5007 line and radial velocity,
as well as the consideration of time, resulted in three objects for
observation. Table\,\ref{sample} presents the coordinates (right ascension
and declination, hereafter RA and Dec, respectively), apparent magnitudes at
[O~{\sc iii}] $\lambda$5007, heliocentric velocities and galactocentric
distances of the three PNe (hereafter named PN1, PN2 and PN3). These
PNe are located in the outer disk of M31, and have been identified as in the
Northern Spur substructure by \citet{mer06}. Their projected galactocentric
distances are from 23 to 27~kpc. Figure\,\ref{m31_pne} shows the spatial
distribution of PNe observed in M31. The sample includes the $\sim$3000 PNe
from \citet{mer06} and those observed by LAMOST\footnote{Large Sky Area
Multi-Object Fiber Spectroscopic Telescope. URL: http://www.lamost.org}
\citep{yuan10}. The coordinates $\xi$ and $\eta$ in Figure\,\ref{m31_pne} are
the offsets in RA and Dec relative to the center of M31, respectively, and
thus define an M31-based reference frame. $\xi$ and $\eta$ are calculated
following the geometric transformations of \citet{hbk91},

\begin{equation}
\label{xi}
\xi = \sin({\rm RA} - {\rm RA}_{0})\cos({\rm Dec}),
\end{equation}
and
\begin{eqnarray}
\label{eta}
\nonumber
\eta = \sin({\rm Dec})\cos({\rm Dec}_{0})\\
 - \cos({\rm RA} - {\rm RA}_{0})\cos({\rm Dec})\sin({\rm Dec}_{0}),
\end{eqnarray}
where RA$_{0}$ = 00$^{\rm h}$42$^{\rm m}$44.4$^{\rm s}$ (J2000.0) and
Dec$_{0}$ = 41$^{\rm o}$16$^{\prime}$08$^{\prime\prime}$ (J2000.0) are the
coordinates of the optical center of M31 and are adopted from \citep{dev91}.
The radial velocities of PN1, PN2 and PN3 are $-$19.6, $-$44.4 and
$-$90.3~km/s, respectively. Thus the corresponding wavelength differences
between the observed [O~{\sc iii}] $\lambda$4363 auroral line and the mercury
line at $\lambda$4358 for the three objects are 4.59, 4.23 and 3.56\,{\AA}.
Given that the width of the mercury line is 5--6\,{\AA} and the full width at
half-maximum (FWHM) of nebular emission line is $\sim$2.40\,{\AA} for the
blue spectra, the [O~{\sc iii}] $\lambda$4363 auroral line is expected to be
resolved from the mercury line for all three objects, provided that
subtraction of the sky background is good.

\subsection{\label{part2:b}
Observations}

We observed three PNe in the Northern Spur outer disk of M31 with the Double
Spectrograph (DBSP) on the Palomar 5.1\,m Hale Telescope. The dichroic D48,
which splits light into separate blue and red channels around 4800\,{\AA} was
used. The two channels cover wavelength ranges 3400--4900\,{\AA} and
4800--7300\,{\AA}. In the observations, the DBSP configuration consists of a
1200 line~mm$^{-1}$ grating blazed at 4150\,{\AA} (with a grating angle of
34.75$^{\rm o}$) in the blue arm and a 316 line~mm$^{-1}$ grating blazed at
6050\,{\AA} (with a grating angle of 23.25$^{\rm o}$) in the red channel.
The blue channel has a thinned, AR coated 2048$\times$4096 (in the 4096
dispersion axis) CCD with 15$\mu$m pixels, and the red channel has a thinned
4096$\times$2048 CCD with 15$\mu$m pixels. Given the seeing condition
($\sim$1.5\,arcsec) at Palomar, the slit width was set to be 1.5\,arcsec;
slit length is 128\,arcsec. This setup enabled us to perform spectroscopy
over the wavelength range 3400--4900\,{\AA} with a resolution of 2.4\,{\AA}
(FWHM) at 0.55\,{\AA}~pixel$^{-1}$, and over the wavelength range
4800--7300\,{\AA} with a resolution of 6.9\,{\AA} (FWHM) at
2.45\,{\AA}~pixel$^{-1}$.

Observations were taken on two nights (Table\ref{sample}). On UT date 2011
September 22, we observed two PNe (PN1 and PN2). The observation
consisted of eight 1800\,s exposures of PN1 and four of PN2 (for both the
blue and red channels of DBSP). The observations were taken under photometric
conditions with a typical seeing of 1.5\,arcsec. On September 23, we observed
PN2 and PN3. The observation consisted of four 1800\,s exposures of PN2 and
six of PN3. The limiting magnitude of the guide star CCD on the telescope
is approximately 16 magnitude, while our target PNe are much fainter.
In order to locate the faint targets in the slit, we adopted the
so-called ``blind offset'' technique: First placed the slit on a selected
bright guide star near each target PN, and then moved the slit to the target
position. In order to avoid light loss due to atmospheric diffraction, the
slit was placed along the parallactic angle during observation. That
was calculated by the telescope control system. The slit was rotated every
30--60 minutes during observation. The data of both nights were calibrated
via a series of exposures of the spectrophotometric standard stars LDS749B
and Hz\,2, which were selected from the ESO Standard Stars
Catalogues\footnote{URL:
http://www.eso.org/sci/observing/tools/standards/spectral.html. The
ultraviolet and optical spectrophotometric data of LDS749B are given by
\citet{oke90}, and those of Hz\,2 are from the unpublished data of J.~B.
Oke.}. Arc lines of an FeAr lamp were used for wavelength calibration for
the blue spectra of the three PNe, and an HeArNe lamp was used for the red.
Exposures of lamps were made before observations of each PN target so that
errors in wavelength calibration can be minimized.

%%%% Table 1. Properties and observations of the three PNe in the Northern
%%%%          Spur of M31:
\begin{table*}
\centering
\caption{Properties and observations of the three Northern Spur PNe.}
\label{sample}
\begin{tabular}{lcccccllr}
\hline
PN ID$^a$ & RA & Dec & $m$($\lambda$5007)$^b$ & $v_{\rm helio}$ & $R_{\rm gal}\,^c$ & \multicolumn{2}{c}{DBSP Exp. (s)} & Slit-width\\
          & (J2000.0) & (J2000.0) &     & (km\,s$^{-1}$)  & (kpc)             & Blue Arm       &  Red Arm         & (arcsec)\\
\hline
PN1 (2426) & 0:47:00.69 & 42:58:55.20 & 20.61 & $-$19.6 & 25.9 & 8$\times$1800 & 8$\times$1800 & 1.5\\
PN2 (2431) & 0:47:58.60 & 43:00:06.48 & 20.83 & $-$44.4 & 27.2 & 8$\times$1800 & 8$\times$1800 & 1.5\\
PN3 (2421) & 0:45:42.60 & 42:55:26.30 & 20.84 & $-$90.3 & 23.9 & 6$\times$1800 & 6$\times$1800 & 1.5\\
\hline
\end{tabular}
\begin{description}
\item[$^a$] Number in the bracket that follows our PN ID is the ID number from
\citet{mer06}.
\item[$^b$] $m$($\lambda$5007) = $-$2.5$\log{F(\lambda5007)} -$ 13.74.
\item[$^c$] Sky-projected galactocentric distance given by \citet{mer06}.
\end{description}
\end{table*}

%%%% Figure 1. Spatial distribution of PNe in M31:

\subsection{\label{part2:c}
Data reduction}

\begin{figure*}
\begin{center}
\includegraphics[width=14cm,angle=0]{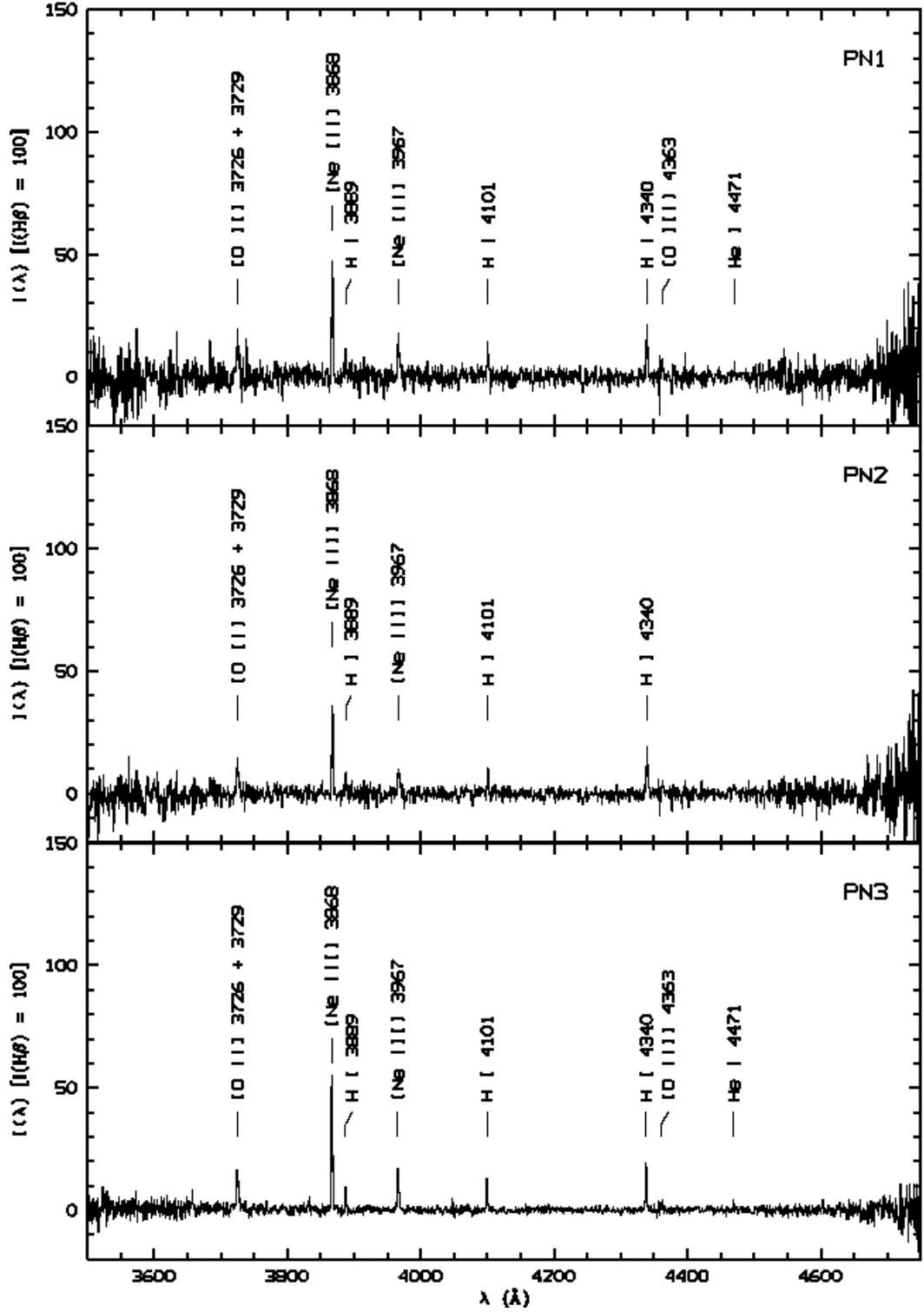}
\caption{The DBSP blue spectra of PN1 ($upper$), PN2 ($middle$) and PN3
($lower$), with identifications of important emission lines labeled. All
spectra have been normalized such that H$\beta$ has an integrated flux of
100. Extinction has been corrected for.}
\label{pne_blue}
\end{center}
\end{figure*}

\begin{figure*}
\begin{center}
\includegraphics[width=14cm,angle=0]{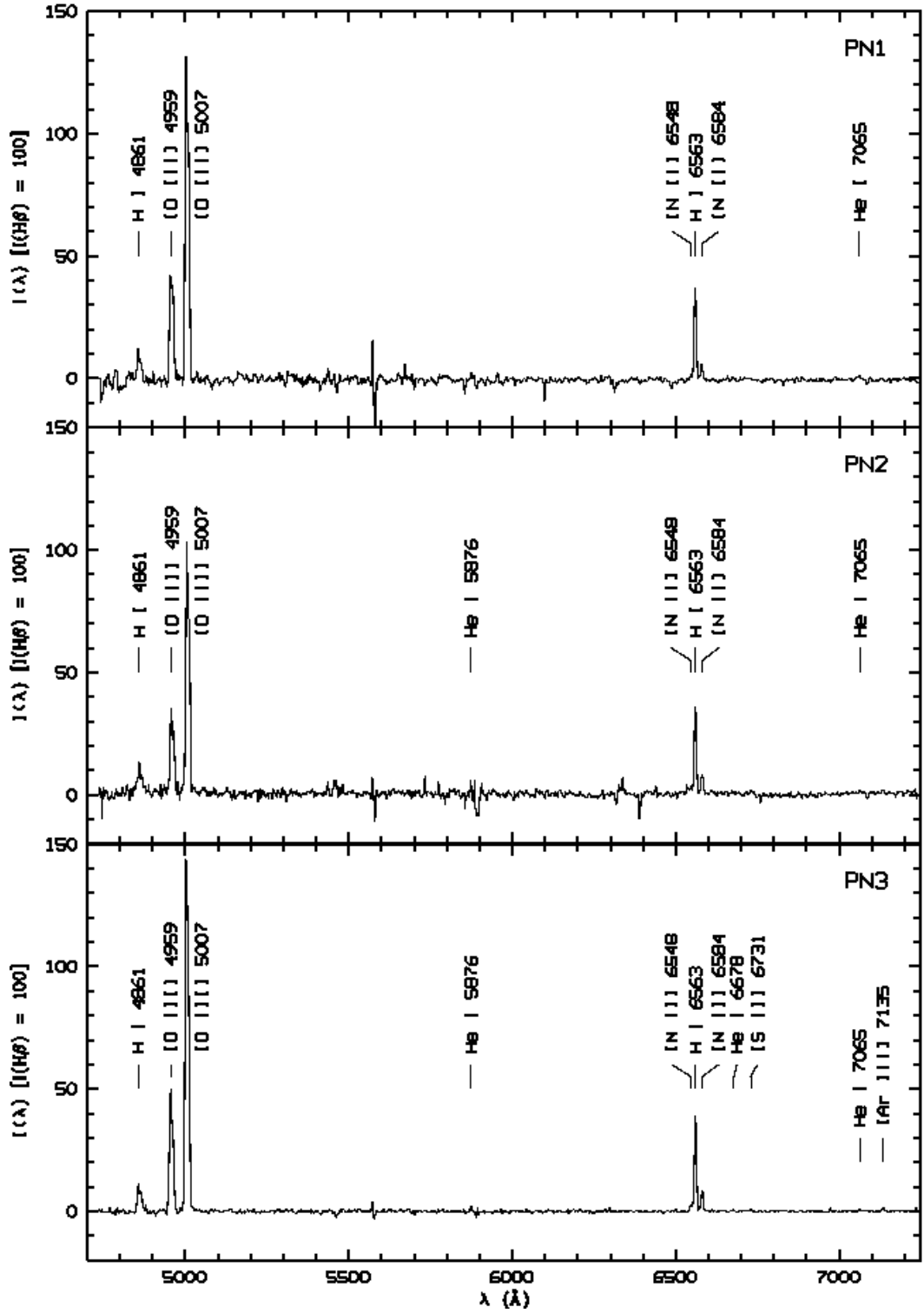}
\caption{Same as Figure\,\ref{pne_blue} but for the DBSP red spectra.}
\label{pne_red}
\end{center}
\end{figure*}

All the data were reduced with standard procedures for long-slit spectra using
the {\sc long92} package in {\sc midas}\footnote{{\sc midas} is developed and
distributed by the European Southern Observatory.}. The raw 2-dimensional
(2-D) spectra were debiased, flat-fielded and cosmic-ray removed, and
then wavelength calibrated using exposures of an FeAr (for the blue data) and
HeArNe (for the red data) lamp. Sky background was then subtracted from
the 2-D spectra. 1-D spectra of the PNe were extracted from the 2-D spectra,
corrected for the atmospheric extinction, and flux calibrated using
observations of the standard stars LDS749B and Hz\,2. The 1-D spectra with
1800\,s exposure for each PN were then combined\footnote{Weighted average of
individual exposures, with assigned weights proportional to the S/N ratios of
the same emission line detected in the blue or red spectrum.}, corrected for
interstellar extinction (see Section\,\ref{part2:d}) and then normalized such
that H$\beta$ has an integrated flux of 100. Here the H$\gamma$ line was used
for the blue spectra and H$\alpha$ for the red, assuming that the
$I$(H$\gamma$)/$I$(H$\beta$) and $I$(H$\alpha$)/$I$(H$\beta$) ratios of H~{\sc
i} are as those calculated by \citet{sh95} at $T_\mathrm{e}$ = 10\,000~K and
$N_\mathrm{e}$ = 10$^{4}$~cm$^{-3}$ in Case~B recombination.
Figures\,\ref{pne_blue} and \ref{pne_red} are the blue and red 1-D spectra of
the three PNe, respectively. The S/N of the blue spectra are systematically
lower than those of the red data, which are mainly due to the relatively low
efficiency of the blue CCD.

\begin{figure*}
\begin{center}
\includegraphics[width=16cm,angle=0]{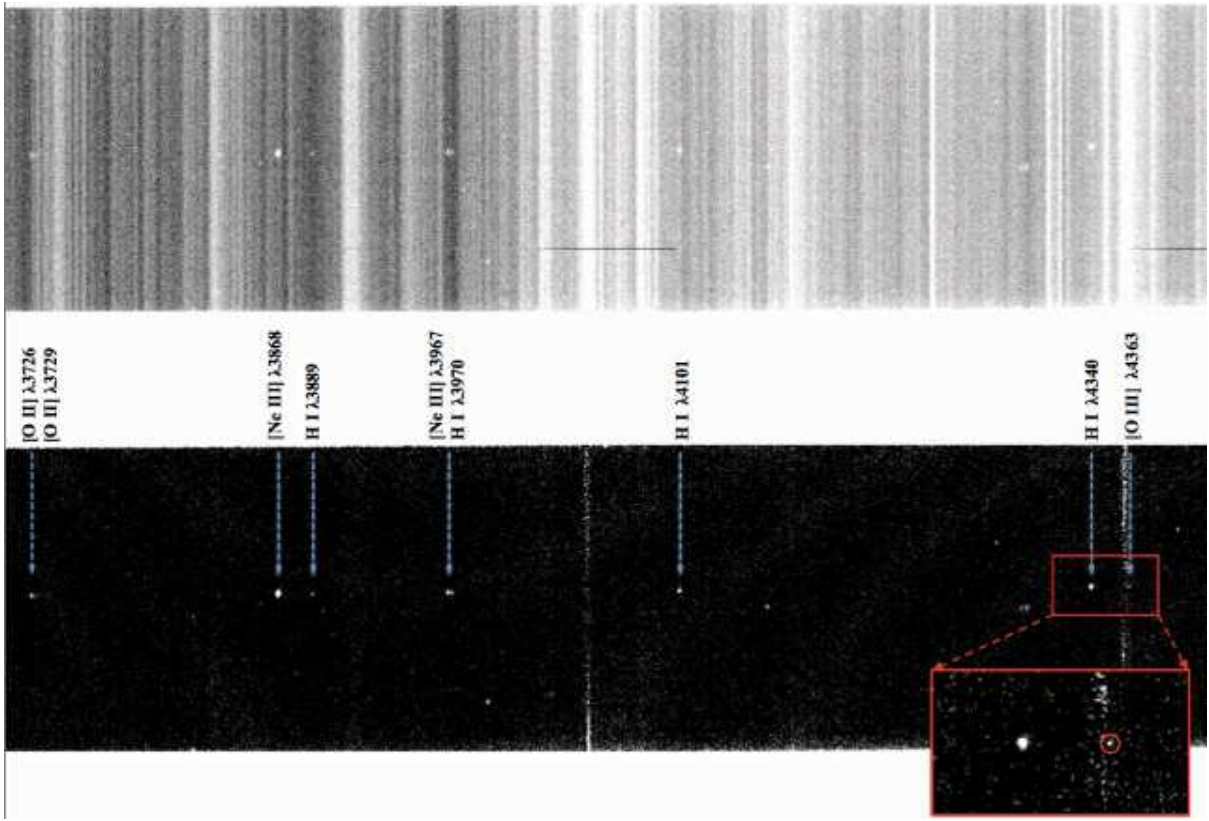}
\caption{2-D blue spectrum of PN1 showing the emission lines detected in the
wavelength range 3710--4405\,{\AA}. The $upper$ and $lower$ panels are the
spectrum before and after subtraction of sky background, respectively. The
two panels are scaled to different levels so that the strong sky
background can be shown in the $upper$ panel while the emission lines of PN1
can be seen in the $lower$ panel. Identifications of the emission lines are
labeled. The weak feature of the [O~{\sc iii}] $\lambda$4363 auroral line is
seen in the sky-subtracted, combined spectrum, as indicated by a red circle
in the inset. Residual from subtraction of the Hg~{\sc i} $\lambda$4358
mercury line is beside the [O~{\sc iii}] $\lambda$4363 line.}
\label{pn1_2d}
\end{center}
\end{figure*}

\begin{figure*}
\begin{center}
\includegraphics[width=16cm,angle=0]{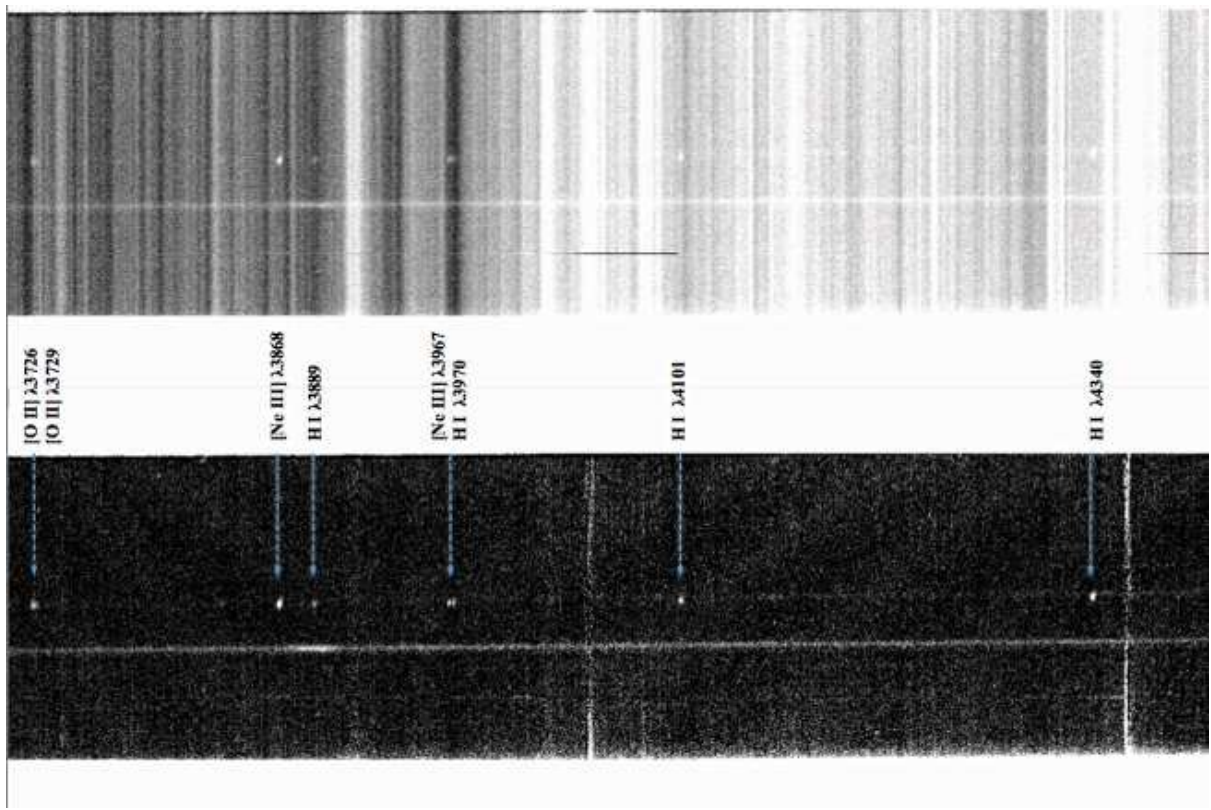}
\caption{Same as Figure\,\ref{pn1_2d} but for PN2. The [O~{\sc iii}]
$\lambda$4363 auroral line is not seen in the sky-subtracted spectrum (the
$lower$ panel). A foreground star is located about 18\,arcsec below
PN2.}
\label{pn2_2d}
\end{center}
\end{figure*}

\begin{figure*}
\begin{center}
\includegraphics[width=16cm,angle=0]{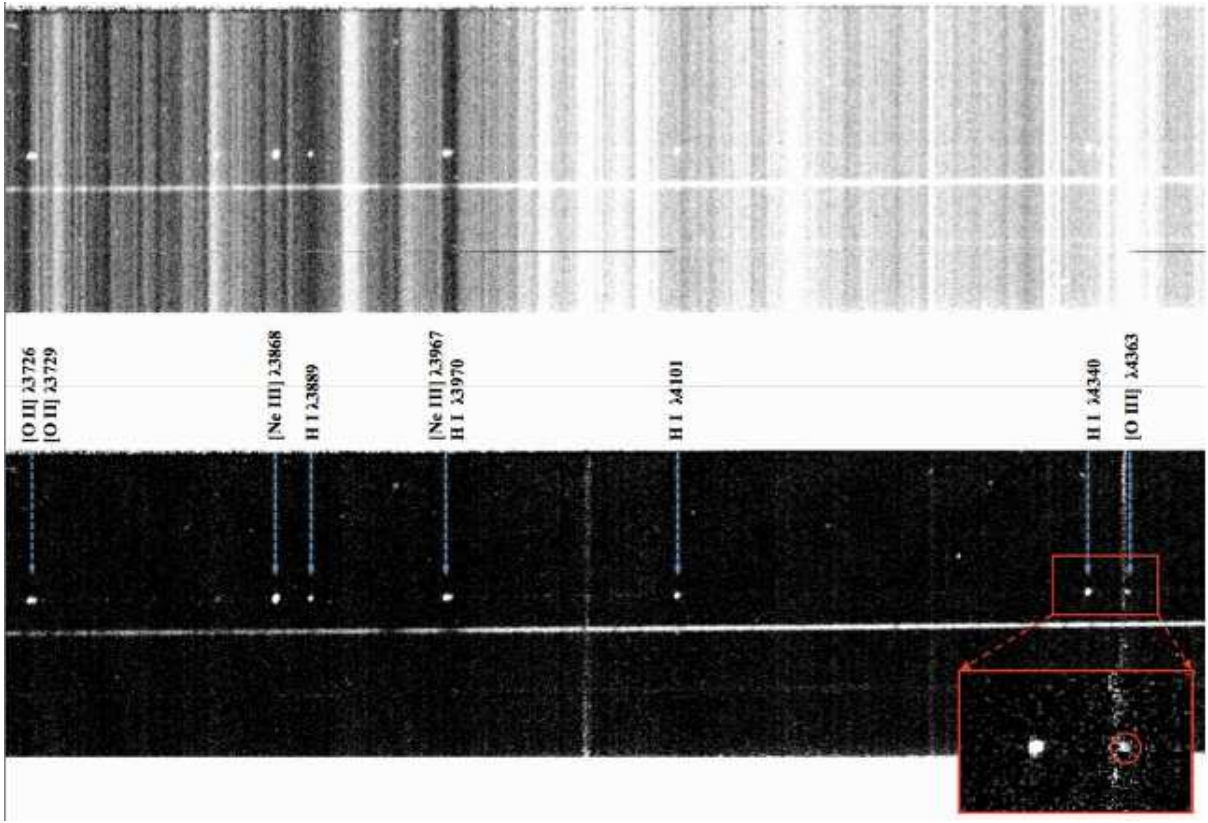}
\caption{Same as Figure\,\ref{pn1_2d} but for PN3. The [O~{\sc iii}]
$\lambda$4363 auroral line is seen in the sky-subtracted spectrum (the $lower$
panel), as indicated by a red circle in the inset. A bright foreground star is
located about 13\,arcsec below PN3.}
\label{pn3_2d}
\end{center}
\end{figure*}

Observations of the three PNe in the Northern Spur of M31 are challenging,
given that the sky background at the Palomar Observatory is strong and our
targets are faint. The very bright mercury line at 4358\,{\AA} makes
it difficult to directly observe the [O~{\sc iii}] $\lambda$4363 auroral line
(Section\,\ref{part2:a}) which can be used to derive the electron
temperature.  We managed to detect emission lines in the three PNe, and
separated the [O~{\sc iii}] $\lambda$4363 line from the mercury line in two
objects. Figures\,\ref{pn1_2d}, \ref{pn2_2d} and \ref{pn3_2d} show the 2-D
blue spectrum of PN1, PN2 and PN3, respectively, in the wavelength region
3710--4405\,{\AA}. The [O~{\sc iii}] $\lambda$4363 auroral line can be seen
in the sky-subtracted spectra of PN1 and PN3. However, residuals
from sky subtraction, especially of the strong mercury line, are still
present in the processed 2-D image (see the lower panels of
Figures\,\ref{pn1_2d}, \ref{pn2_2d} and \ref{pn3_2d}). That is because
sky subtraction was done by choosing two separate sections on the slit, one
on either side of a target spectrum, with a separation of about 3--5 pixels
from the PN, so that the effects of image distortion along the slit can be
minimized.

Consideration has been given to the distortion of emission lines. We
tried to map out the distortion of the Hg~{\sc i} $\lambda$4358 mercury line
along the slit. That was done by measuring the peak wavelengths of the mercury
line in the wavelength-calibrated 2-D image along the slit direction, with
every five rows of CCD pixels binned together.  The peak positions of the
mercury line were well fitted with linear regression.
%as shown in Figure\,\ref{distortion}.
The linear fit yields a very small angle relative to the slit direction, which
indicates that the distortion is small. We also checked the distribution of
the mercury line flux along the slit, and found that is mostly homogeneous.
Besides, the sky background regions defined on the slit (two regions on the
slit, with one on either side of a target PN) are very close to the target
PN. Thus the effects of distortion of the sky lines on the PN emission line
measurements are negligible. Nevertheless, we applied the distortion
correction to all 2-D frames. Despite much effort has been paid, not much
can be further improved in sky subtraction. Given the relatively strong sky
background compared to the intrinsic faintness of our target PNe, it is
difficult to completely remove the sky background over the whole slit on the
2-D frame.

Since the PNe in M31 are point sources, emission lines on a 2-D spectrum
are detected as bright spots (see Figures\,\ref{pn1_2d}, \ref{pn2_2d} and
\ref{pn3_2d}) with 5-10 pixels in diameter depending on strength
of the line. The dispersion direction of the blue spectrum is not always
perpendicular to the slit. We carried out polynomial fits to the
positions (wavelength in {\AA} versus the row number of CCD along the slit)
of emission lines on the wavelength-calibrated and sky-subtracted 2-D frame,
and then extracted the 1-D spectrum by averaging a number of pixels in the
direction of slit (about 20 pixels with the target PN in the center) along
the track of the polynomial fit. Although residuals from sky subtraction are
present in the 2-D image, they are only the strongest far away from the
target PN and negligible in the region close by, as seen in
Figures\,\ref{pn1_2d} and \ref{pn3_2d}. Thus measurements of the [O~{\sc
iii}] $\lambda$4363 line are not much affected. The dispersion direction
of the red spectra is well perpendicular to the slit, and we averaged the
CCD rows along the slit directly to get 1-D spectra.
The spectra of the two standards (LDS749B and Hz\,2) were used to
derive the response curves of the blue and red channels for dichroic D48.
The efficiency is very low above 4700\,{\AA} in the blue channel and below
4800\,{\AA} in the red. The extinction-corrected 1-D blue spectra has not
enough quality above 4700\,{\AA} as seen in Figure\,\ref{pne_blue}, and the
H$\beta$ line can not be used to join the blue and red spectra as expected.
Given the relatively low efficiency near 4800\,{\AA} on the red CCD, we used
the H$\alpha$ line instead of H$\beta$ to normalize line fluxes such that
$I$(H$\alpha$) = 285. For the blue spectra, all line fluxes were scaled such
that $I$(H$\gamma$) = 47. The normalization is based on the theoretical
H~{\sc i} line ratios and the assumption that $I$(H$\beta$) = 100.

%%%% Figure 2. Blue spectra of the three NS PNe:

%%%% Figure 3. Red spectra of the three NS PNe:

%%%% Figure 4. 2-D spectrum of PN1:

%%%% Figure 5. 2-D spectrum of PN2:

%%%% Figure 6. 2-D spectrum of PN3:

%%%% Figure 7. Distortion of the mercury line at 4358 Angstrom along the slit:
%\begin{figure*}
%\begin{center}
%\includegraphics[width=10cm,angle=0]{fig7.eps}
%\caption{Distortion of the mercury line Hg~{\sc i} $\lambda$4358 along the
%slit. The slit direction is along the horizontal axis and the dispersion
%direction is parallel to the vertical. Central wavelengths and half widths of
%the mercury line are represented by black circles and bars,
%respectively; each data point is binned from five pixels along the slit. The
%red line is a linear regression fit to the positions of the mercury line
%peaks, which yields a position angle of 0.15$^{\rm o}$ with respect to the
%slit. Slit position (horizontal axis) is centered on the [O~{\sc iii}]
%$\lambda$4363 auroral line (the black filled circle). The observed wavelengths
%of the mercury line peaks (vertical axis) are relative to that of the [O~{\sc
%iii}] line, $\Delta\lambda$ = $\lambda_{\rm obs}$(Hg~{\sc i} $\lambda$4358)
%$-$ $\lambda_{\rm obs}$([O~{\sc iii}] $\lambda$4363). The sky background
%defined on the slit is indicated by the two grey regions, one on either side
%of a target PN. This figure is based on the data reduction of PN1.}
%\label{distortion}
%\end{center}
%\end{figure*}

\subsection{\label{part2:d}
Extinction correction}

The logarithmic extinction parameter at H$\beta$, $c$(H$\beta$), was
derived for the three M31 PNe by comparing the observed H~{\sc i} Balmer line
ratio, $I$(H$\gamma$)/$I$(H$\delta$), with the predicted Case~B value. That
yields a $c$(H$\beta$) value of 0.43, 0.51 and 0.54 for PN1, PN2 and PN3,
respectively.
The H~{\sc i} $\lambda$3889 ($n$ = 2\,--\,8) and $\lambda$3970 ($n$ =
2\,--\,7) lines were not used for extinction correction because the former
line is blended with He~{\sc i} $\lambda$3888 (2s\,$^{3}$S --
3p\,$^{3}$P$^{\rm o}$) and the latter is blended with the [Ne~{\sc iii}]
$\lambda$3967 (2p$^4$~$^3$P$_{1}$ -- $^1$D$_{2}$) line. The observed line
fluxes were dereddened by

\begin{equation}
\label{deredden}
I(\lambda) = 10^{c(\rm{H}\beta) f(\lambda)} F(\lambda),
\end{equation}
where $f(\lambda)$ is the extinction curve (relative to the $f$ value at
H$\beta$) adopted from \citet{ccm89} with a total-to-selective
extinction ratio $R_{\rm V}$ = 3.1.

%The extinction parameter $c$(H$\beta$) derived for the three PNe are
%all higher than those given by \citet{kwi12}, who observed 16 PNe in the
%outer disk of M31 with galactocentric distances of 18--43~kpc, but generally
%agrees the observations of three M31 disk PNe by \citet{jc99}. The
%extinction derived by Jacoby ...

The method of extinction correction described above assumes that the Galactic
foreground extinction is negligible compared to the local extinction in M31.
In order to assess how important the Galactic foreground extinction can be to
the line fluxes, we carried out another reddening correction procedure: first
correct for the MW foreground extinction, and then correct for the
local extinction. The foreground reddening $E$($B\,-\,V$) toward M31 is
0.062, which was adopted from \citet{sch98} who conducted an all-sky survey
of infrared dust emission and used this to calculate the reddening
$E$($B\,-\,V$) across the sky to within an uncertainty of $\sim$16 per cent.
After the MW foreground extinction toward M31 had been corrected for,
we used the H~{\sc i} $I$(H$\alpha$)/$I$(H$\beta$) ratio to corrected for the
local extinction. The derived logarithmic extinction parameter $c$(H$\beta$)
of PN1, PN2 and PN3 are 0.30, 0.31 and 0.45, respectively.

The extinction-corrected line fluxes based on the above two methods differ
by $\sim$10 per cent. Since correction for the MW foreground extinction may
introduce extra uncertainties and that so far most observations of the M31
nebulae (PNe and H~{\sc ii} regions) only corrected for extinction once
(e.g. \citealt{jc99}; \citealt{kwi12}; \citealt{san12}; \citealt{zb12})
using a certain extinction law, we adopted the first set of the extinction
parameters in our data analysis (Table\,\ref{lines}). If the same
extinction law (e.g. \citealt{ccm89}) is used, we would expect that the MW
and M31 extinctions could be ``packed'' in to a single $c$(H$\beta$), which
is equivalent to the first method.

\section{\label{part3}
Results}

\subsection{\label{part3:a}
Relative line intensities}

\begin{figure*}
\begin{center}
\includegraphics[width=12cm,angle=0]{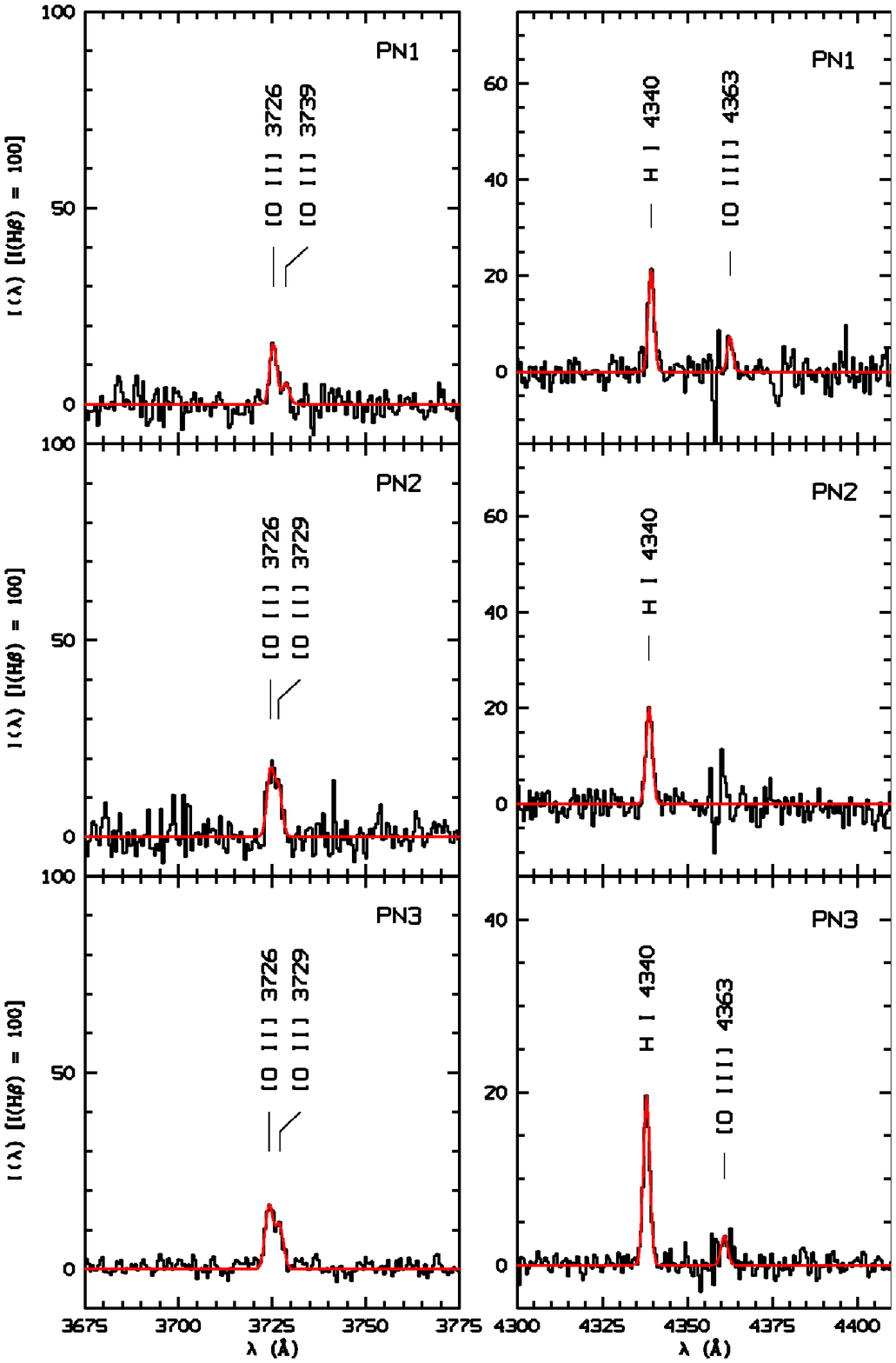}
\caption{The blue spectra of PN1 ($top$), PN2 ($middle$) and PN3 ($bottom$),
showing the observed [O~{\sc ii}] and [O~{\sc iii}] lines. The red continuous
curves are Gaussian-profile fits to the [O~{\sc ii}] $\lambda\lambda$3726 and
3729 doublet (the $left$ column) and the [O~{\sc iii}] $\lambda$4363 line
(the $right$ column). %Line fitting is not given for the [O~{\sc ii}] doublet
%of PN1 due to relatively poor S/N (the $upper$-$left$ panel).
The [O~{\sc iii}] $\lambda$4363 line is not detected in PN2, and the feature
at 4360\,{\AA} in its blue spectrum (the $middle$-$right$ panel) is likely to
be the residual of sky subtraction. Spectra have been normalized such that
H$\beta$ has an integrated flux of 100. Extinction has been corrected for.}
\label{fits}
\end{center}
\end{figure*}

The extinction-corrected relative line intensities of the three PNe are
presented in Table\,\ref{lines}. As mentioned in Section\,\ref{part2:c}, the
H$\gamma$ line was used for flux normalization of the blue data and H$\alpha$
was used for the red. Here the Balmer line ratios $I$(H$\gamma$)/$I$(H$\beta$)
and $I$(H$\alpha$)/$I$(H$\beta$) were adopted from the Case~B theoretical
calculations. The assumed electron temperature and density are valid because
the H~{\sc i} Balmer line ratios are mostly insensitive to these two
physical quantities.
%According to the calculations of \citet{sh95}, the
%$I$(H$\alpha$)/$I$(H$\beta$) and $I$(H$\gamma$)/$I$(H$\beta$) ratios of H~{\sc
%i} only differ by 10 and 4 per cent, respectively, when the temperature and
%density change from one extreme case, $T_\mathrm{e}$ = 5000~K and
%$N_\mathrm{e}$ = 10$^{2}$~cm$^{-3}$, to another one, $T_\mathrm{e}$ =
%20\,000~K and $N_\mathrm{e}$ = 10$^{6}$~cm$^{-3}$.
Continua were not clearly detected in the spectra due to faintness of the
objects. Thus the integrated line fluxes were directly obtained from Gaussian
profile fits. As an example, Figure\,\ref{fits} shows Gaussian profile fits
to the [O~{\sc ii}] and [O~{\sc iii}] lines detected in the blue spectra of
the three PNe. Uncertainties in line intensities were estimated from direct
integration and the Gaussian profile fitting of the profile of each
emission line. PN3 has the relatively higher S/N in the blue spectra
among the three PNe. For example, the S/N ratio of the H$\gamma$
$\lambda$4340 line in the blue spectrum of PN3 is 23.4, while those ratios of
of PN1 and PN2 are 9.8 and 9.5, respectively. Differences in data quality of
the blue spectra of the three PNe can be seen in Figure\,\ref{pne_blue}.

Although the Hg~{\sc i} $\lambda$4358 mercury line at the Palomar Observatory
is bright, it was expected that the [O~{\sc iii}] $\lambda$4363 line could
be detected in all the three PNe, given the radial velocities of the three
objects (Table\,\ref{sample}), if subtraction of the sky background is well
done. However, the [O~{\sc iii}] $\lambda$4363 line was not detected in the
blue spectrum of PN2 (Figures\,\ref{pn2_2d} and \ref{fits}), which has a
medium radial velocity ($-$44.4\,km/s) among the three objects, probably due
to the relatively low S/N ratios. The [O~{\sc iii}] $\lambda$4363 line was
detected in PN1 (Figures\,\ref{pn1_2d} and \ref{fits}), which has the lowest
radial velocity ($-$19.6\,km/s), with a S/N ratio of 5.3. The S/N ratios of
the blue spectrum of PN3 are higher than the other two objects, thus
consequently, although the radial velocity of PN3 is the highest among the
three ($-$90.3\,km/s), a weak feature of the [O~{\sc iii}] $\lambda$4363 line
was still clearly detected (Figures\,\ref{pn3_2d} and \ref{fits}), with a S/N
ratio of $\sim$4.0, which is close to that of PN1. Although much effort has
been made to subtract the sky background (see Section\,\ref{part2:c}),
accurate measurements of the [O~{\sc iii}] $\lambda$4363 line for our PN
sample are still affected by the residual of sky subtraction, which can be
seen in Figures\,\ref{pn1_2d} and \ref{pn3_2d}. Integrated fluxes of the
[O~{\sc iii}] $\lambda$4363 line of PN1 and PN3 are given in
Table\,\ref{lines}, and the uncertainties in fluxes were estimated from
Gaussian profile fitting (Figure\,\ref{fits}). Electron temperature
diagnostics for PN1 and PN3 using the [O~{\sc iii}]
($\lambda$4959\,+\,$\lambda$5007)/$\lambda$4363 line ratio are presented in
Section\,\ref{part3:b}.

The [O~{\sc ii}] $\lambda\lambda$3726,\,3729 lines are detected in
the blue spectrum of PN3. Two Gaussian profiles were used to fit the doublet
(Figure\,\ref{fits}), which yields an $I$($\lambda$3726)/$I$($\lambda$3729)
intensity ratio of 1.4. Measurement uncertainties of the two [O~{\sc
ii}] lines for PN3 are $\sim$20--30 per cent. The double-peak profile of the
[O~{\sc ii}] doublet lines in the spectrum of PN2 are not so obvious as in
PN3 whose data quality are better. However, Gaussian profile fitting yields
an $I$($\lambda$3726)/$I$($\lambda$3729) ratio of 1.5 for PN2, which
is quite reasonable. Uncertainties in the line fluxes of the [O~{\sc ii}]
lines of PN2 could be $\sim$30--40 per cent. The [O~{\sc ii}] line
ratio observed in PN1 2.8.  Electron densities are estimated for the three
PNe using the [O~{\sc ii}] line ratio in Section\,\ref{part3:b}.

Some important emission lines detected suffer from blending. The [Ne~{\sc
iii}] $\lambda$3967 line is blended with H~{\sc i} $\lambda$3970. We
corrected for the flux contribution from the H~{\sc i} $\lambda$3970 line
using the theoretical H~{\sc i} $I$($\lambda$3970)/$I$($\lambda$4340) ratio.
The Ne$^{2+}$/H$^{+}$ abundances were then derived from the corrected [Ne~{\sc
iii}] $\lambda$3967 line flux. The H~{\sc i} $\lambda$3889 line is blended
with the He~{\sc i} 2s\,$^{3}$S -- 3p\,$^{3}$P$^{\rm o}$ $\lambda$3888 line
(Table\,\ref{lines}). We corrected the He~{\sc i} line flux for the
contribution from the H~{\sc i} $\lambda$3889 line, using the observed flux
of the H~{\sc i} $\lambda$4340 line and the theoretical H~{\sc i}
$I$($\lambda$3889)/$I$($\lambda$4340) ratio. The corrected fluxes of the
He~{\sc i} $\lambda$3888 line were used to derived the He$^{+}$/H$^{+}$
abundance for the three PNe (Section\,\ref{part3:c}). The [S~{\sc ii}]
$\lambda$6731 (3p$^{3}$~$^{4}$S$^{\rm o}_{3/2}$ -- $^{2}$D$^{\rm o}_{3/2}$)
line is detected in the red spectrum of PN3 (Figure\,\ref{pne_red}), with an
S/N ratio of $\sim$3--4. The other component of the [S~{\sc ii}] doublet,
the $\lambda$6716 (3p$^{3}$~$^{4}$S$^{\rm o}_{3/2}$ -- $^{2}$D$^{\rm o}_{5/2}$)
line, which is expected to be weaker than $\lambda$6731, is not clearly seen
due to weakness. The [Ar~{\sc iii}] $\lambda$7136 (3p$^{4}$~$^{3}$P$_{2}$ --
$^{1}$D$_{2}$) line is also detected in the spectrum of PN3. The
S$^{+}$/H$^{+}$ and Ar$^{2+}$/H$^{+}$ ionic abundances are derived from the
$\lambda$6731 and $\lambda$7136 lines, respectively, in
Section\,\ref{part3:c}.

%%%% Table 2. Emissoin line table:
\begin{table*}
\caption{Emission lines detected in the spectra of the three Northern
Spur PNe. All intensities have been corrected for extinction. For the red
spectra (3400--4800\,{\AA}), line intensities are normalized such that
$I$(H$\gamma$) = 47; for the red spectra (4700--7300\,{\AA}), line intensities
are normalized such that $I$(H$\alpha$) = 285.
%An asterisk `` * '' denotes the emission line of that PN is blended with an
%adjacent feature.
A colon `` : '' denotes the uncertainty in line flux is large ($>$100\%).}
\label{lines}
\centering
\begin{tabular}{lclccc}
\hline
Ion & $\lambda$ & Transition & \multicolumn{3}{c}{$I(\lambda)$}\\
 & ({\AA}) &    &  PN1 & PN2 & PN3\\
\hline
$[$O~{\sc ii}$]$   & 3726 & 2p$^3$~$^4$S$^{\rm o}_{3/2}$ -- 2p$^3$~$^2$D$^{\rm o}_{3/2}$ &   50$\pm$10  &   58$\pm$14  &   43$\pm$8  \\
$[$O~{\sc ii}$]$   & 3729 & 2p$^3$~$^4$S$^{\rm o}_{3/2}$ -- 2p$^3$~$^2$D$^{\rm o}_{5/2}$ & 17.8$\pm$6.5 &   42$\pm$12  &   29$\pm$6  \\
$[$Ne~{\sc iii}$]$ & 3868 & 2p$^4$~$^3$P$_{2}$ -- 2p$^4$~$^1$D$_{2}$                     &  123$\pm$11  &   94$\pm$10  &  136$\pm$5  \\
 H~{\sc i}         & 3889 & 2p~$^2$P$^{\rm o}$ -- 8d~$^2$D                               &   24$\pm$3   &   21$\pm$4   &   21$\pm$3  \\
% He~{\sc i}        & 3888 & 2s~$^3$S -- 3p~$^3$P$^{\rm o}$                               &         *    &         *    &         *   \\
$[$Ne~{\sc iii}$]$ & 3967 & 2p$^4$~$^3$P$_{1}$ -- 2p$^4$~$^1$D$_{2}$                     &   53$\pm$9   &   45$\pm$6   &   56$\pm$4  \\
% H~{\sc i}         & 3970 & 2p~$^2$P$^{\rm o}$ -- 7d~$^2$D                               &         *    &         *    &         *   \\
 H~{\sc i}         & 4101 & 2p~$^2$P$^{\rm o}$ -- 6d~$^2$D                               &   30$\pm$5   &   35$\pm$7   &   25$\pm$2  \\
 H~{\sc i}         & 4340 & 2p~$^2$P$^{\rm o}$ -- 5d~$^2$D                               &   47$\pm$8   &   47$\pm$7   &   47$\pm$4  \\
$[$O~{\sc iii}$]$  & 4363 & 2p$^2$~$^1$D$_{2}$ -- 2p$^2$~$^1$S$_{1}$                     & 14.5$\pm$4.4 &              &  8.3$\pm$2.5\\
 He~{\sc i}        & 4471 & 4d~$^3$D -- 2p~$^3$P$^{\rm o}$                               &  8.8$\pm$2.9 &              &  7.1$\pm$2.2\\
 H~{\sc i}$^a$     & 4861 & 2p~$^2$P$^{\rm o}$ -- 4d~$^2$D                               &       100    &      100     &      100    \\
$[$O~{\sc iii}$]$  & 4959 & 2p$^2$~$^3$P$_{1}$ -- 2p$^2$~$^1$D$_{2}$                     &  406$\pm$24  &  302$\pm$20  &  460$\pm$7  \\
$[$O~{\sc iii}$]$  & 5007 & 2p$^2$~$^3$P$_{2}$ -- 2p$^2$~$^1$D$_{2}$                     & 1075$\pm$19  &  875$\pm$18  & 1238$\pm$6  \\
 He~{\sc i}        & 5876 & 2p~$^3$P$^{\rm o}$ -- 3d~$^3$D                               &              & 12.8$\pm$3.9 & 11.9$\pm$2.3\\
$[$N~{\sc ii}$]$   & 6548 & 2p$^2$~$^3$P$_{1}$ -- 2p$^2$~$^1$D$_{2}$                     &   11$\pm$3   &   31$\pm$9   &   25$\pm$5  \\
 H~{\sc i}         & 6563 & 2p~$^2$P$^{\rm o}$ -- 3d~$^2$D                               &  285$\pm$7   &  285$\pm$8   &  285$\pm$2  \\
$[$N~{\sc ii}$]$   & 6583 & 2p$^2$~$^3$P$_{2}$ -- 2p$^2$~$^1$D$_{2}$                     &   44$\pm$7   &   77$\pm$10  &   65$\pm$3  \\
 He~{\sc i}        & 6678 & 2p~$^1$P$^{\rm o}$ -- 3d~$^1$D                               &              &              &  5.0$\pm$2.7\\
$[$S~{\sc ii}$]$   & 6731 & 2p$^3$~$^4$S$^{\rm o}_{3/2}$ -- 2p$^3$~$^2$D$^{\rm o}_{3/2}$ &              &              &  6.1$\pm$2.9\\
 He~{\sc i}        & 7065 & 2p~$^3$P$^{\rm o}$ -- 3s~$^3$S                               &  7.8$\pm$:   &  12.1$\pm$:  &  6.2$\pm$:  \\
$[$Ar~{\sc iii}$]$ & 7136 & 3p$^4$~$^3$P$_{2}$ -- 3p$^4$~$^1$D$_{2}$                     &              &              &   13$\pm$3  \\
\\
$c$(H$\beta$) &    &                                                                     &      0.43    &       0.51   &      0.54   \\
\hline
\end{tabular}
\begin{description}
\item[$^a$] The integrated flux of H$\beta$ is assumed to be 100.
\end{description}
\end{table*}

%%%% Figure 7. Gaussian profile fits to the [O II] and [O III] lines:

\subsection{\label{part3:b}
Plasma diagnostics}

Plasma diagnostics were carried out using the collisionally excited lines
(CELs) of heavy elements (e.g. \citealt{of06}) detected in the spectra of our
PN sample. The [O~{\sc ii}] $\lambda$3726/$\lambda$3729 line ratio was used
to determine the electron density, and the [O~{\sc iii}]
($\lambda$4959\,+\,$\lambda$5007)/$\lambda$4363 nebular-to-auroral line ratio
was used to determine the electron temperature. Results of plasma diagnostics
are given in Table\,\ref{te_ne}.

\begin{figure*}
\begin{center}
\includegraphics[width=5.25cm,angle=-90]{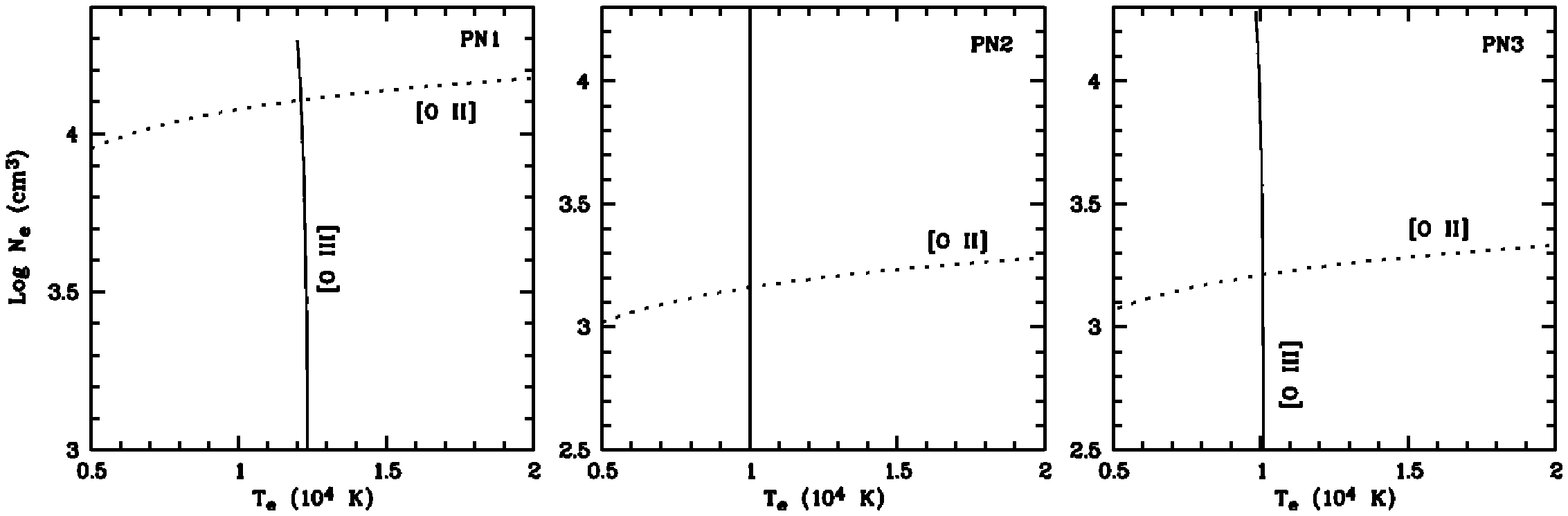}
\caption{Plasma diagnostic diagrams for PN1 ($left$), PN2 ($middle$) and PN3
($right$). The solid curve is the temperature diagnostic using the [O~{\sc
iii}] ($\lambda$4959\,+\,$\lambda$5007)/$\lambda$4363 ratio; the dotted curve
is the density diagnostic using the [O~{\sc ii}] $\lambda$3726/$\lambda$3729
ratio. The electron temperature of PN2 is assumed to be 10\,000~K because the
[O~{\sc iii}] $\lambda$4363 line is not detected in the spectrum of this
object.}
\label{diagnostics}
\end{center}
\end{figure*}

Figure\,\ref{diagnostics} shows the plasma diagnostic diagrams of the three
PNe. They were created by solving the level population equations for five-level
atomic models using the program {\sc equib}\footnote{{\sc equib} was originally
developed by I.~D. Howarth et al. from the Department of Physics and Astronomy,
University College London, for calculating the level population equations of
multi-level ($n\geq$5) atomic models.}. The [O~{\sc iii}]
($\lambda$4959\,+\,$\lambda$5007)/$\lambda$4363 nebular-to-auroral and the
[O~{\sc ii}] $\lambda$3726/$\lambda$3729 nebular line ratios observed in PN1
yield an electron temperature of 12\,200~K and a density of
$\sim$12\,800~cm$^{-3}$, respectively. The [O~{\sc iii}] and [O~{\sc ii}]
line ratios observed in the spectrum of PN3 yield an electron temperature of
10\,080~K and a density of $\sim$1650~cm$^{-3}$, respectively. Since the
[O~{\sc iii}] $\lambda$4363 line was not detected in the spectrum of PN2
(Figure\,\ref{fits}), an electron temperature of 10$^{4}$~K, a typical
forbidden-line temperature of PNe, was assumed for this object. At this
temperature, an electron density of $\sim$1450~cm$^{-3}$ was derived for PN2
from the observed [O~{\sc ii}] line ratio.

The derived electron density of PN1 is significantly higher than
those of PN2 and PN3 and its temperature slightly higher, which indicates PN1
may be a relatively younger (compact) PN compared to the other two objects.
However, the density derived for PN1 could be of relatively large uncertainty,
given the faintness of the [O~{\sc ii}] $\lambda$3729 line. In order to check
how much the electron density would affect the resultant ionic abundances, we
carried out abundance calculations at two density cases, $N_\mathrm{e}$ =
10$^{3}$ and 10$^{4}$~cm$^{-3}$, at $T_\mathrm{e}$ = 12\,200~K. The ionic
abundances derived from the [O~{\sc iii}] and [Ne~{\sc iii}] CELs at the two
density cases differ by only 3 per cent, and the ionic abundances derived
from the [N~{\sc ii}] nebular lines differ by about 10 per cent at these two
densities; that difference increases to nearly 40 per cent for the
O$^{+}$/H$^{+}$ ionic abundances derived from the [O~{\sc ii}] nebular lines.
The above differences in the ionic abundances at the two densities are
expected because of the large differences in critical
densities\footnote{Critical densities are all quoted for an electron
temperature of 10$^{4}$~K.}: The critical densities of the [O~{\sc iii}] and
[Ne~{\sc iii}] nebular lines (which have the same upper level $^{1}$D$_{2}$
for each ion) are 6.8$\times$10$^{5}$ and 9.5$\times$10$^{6}$~cm$^{6}$,
respectively, which are much higher than the typical PN densities; the
critical density of the [N~{\sc ii}] $\lambda\lambda$6548, 6583 nebular
lines (which also have the same upper level $^{1}$D$_{2}$) is
6.6$\times$10$^{4}$~cm$^{-3}$, which is also relatively higher than the
average PN density; the critical densities of the [O~{\sc ii}]
$\lambda\lambda$3726, 3729 nebular lines (whose upper levels are the
$^{2}$D$^{\rm o}_{3/2}$ and $^{2}$D$^{\rm o}_{5/2}$ fine-structure levels,
respectively) are 1.5$\times$10$^{4}$ and 3400~cm$^{-3}$, respectively, which
are comparable to the typical PN densities. It can be seen from the diagnostic
diagrams of PN1 and PN3 (Figure\,\ref{diagnostics}) that the [O~{\sc iii}]
line ratio is very sensitive to temperature, and variation in electron density
does not change the resultant temperature much.

%%%% Table 3. Plasma diagnostics for the three NS PNe:
\begin{table*}
\caption{Plasma diagnostics.}
\label{te_ne}
\centering
\begin{tabular}{llll}
\hline
Diagnostic ratio &   PN1   &   PN2   &   PN3\\
\hline
 & \multicolumn{3}{c}{$T_\mathrm{e}$ (K)}\\
$[$O~{\sc iii}$]$ ($\lambda$4959\,+\,$\lambda$5007)/$\lambda$4363 & 12\,200$\pm$1200 & 10\,000$^a$ & 10\,080$\pm$800\\
\\
           & \multicolumn{3}{c}{$N_\mathrm{e}$ (cm$^{-3}$)}\\
$[$O~{\sc ii}$]$ $\lambda$3726/$\lambda$3729  & 12\,800$\pm$3800 & 1450$\pm$500 & 1650$\pm$300\\
\hline
\end{tabular}
\begin{description}
\item [$^a$] An assumed electron temperature for PN2 because its [O~{\sc iii}]
$\lambda$4363 line is not observed.
\end{description}
\end{table*}

%%%% Figure 8. Plasma diagnostic diagrams:

\subsection{\label{part3:c}
Ionic and elemental abundances}

Ionic abundances derived from the emission lines detected in the spectra of
the three PNe are presented in Table\,\ref{ionic}. In order to derive ionic
abundances from CELs, the equations of statistical equilibrium were solved
using the program {\sc equib} to derive the population of the upper level of
a transition. The electron temperatures and densities given in
Table\,\ref{te_ne}, which were yielded from plasma diagnostics
(Section\,\ref{part3:b}), were assumed for abundance determinations.

Several He~{\sc i} lines are detected. The $\lambda$3888 line is
blended with H~{\sc i} H8 $\lambda$3889, whose flux contribution was estimated
from the observed H~{\sc i} $\lambda$4340 line and the theoretical H~{\sc i}
$I$($\lambda$3889)/$I$($\lambda$4340) ratio at $T_\mathrm{e}$ = 10\,000~K and
$N_\mathrm{e}$ = 10$^{4}$~cm$^{-3}$ in Case~B. The flux corrected He~{\sc i}
$\lambda$3888 line yields He$^{+}$/H$^{+}$ ionic abundances that are generally
lower than the other He~{\sc i} lines, $\lambda\lambda$4471, 5876 and 6678,
which are too weak to be accurately measured. However, the He$^{+}$/H$^{+}$
abundance ratio derived from the $\lambda$5876 line observed in PN3 agrees
with that derived from the $\lambda$3888 line. The effective recombination
coefficients for the He~{\sc i} spectrum calculated by \citet{bss99} were
used for the abundance determinations. The $\lambda\lambda$3888 and 5876
lines are among the strongest He~{\sc i} optical recombination lines observed
in PNe. At $T_\mathrm{e}$ = 10\,000~K and $N_\mathrm{e}$ = 10$^{4}$~cm$^{-3}$,
the $I$($\lambda$3888)/$I$($\lambda$4471) and
$I$($\lambda$5876)/$I$($\lambda$4471) ratios of He~{\sc i} are 2.6
and 2.9, respectively, while the intensities of the other He~{\sc i}
optical recombination lines relative to $\lambda$4471 are all less than 0.95
\citep{bss99}. That pattern of the relative intensities of the He~{\sc i}
optical lines remains when the physical condition changes from $T_\mathrm{e}$
= 5000~K and $N_\mathrm{e}$ = 10$^{2}$~cm$^{-3}$ to $T_\mathrm{e}$ = 20\,000~K
and $N_\mathrm{e}$ = 10$^{6}$~cm$^{-3}$. The reason that we chose
$T_\mathrm{e}$ = 5000 and 20\,000~K as the lower and upper limits of the
physical condition is that so far electron temperatures of PNe derived from
the He~{\sc i} recombination line ratios mostly lie in the range
5000--12\,000~K (e.g. \citealt{zhang05}). Given that the H~{\sc i} Balmer
line ratios are mostly insensitive to temperature and density, the line flux
of He~{\sc i} $\lambda$3888 corrected for the contribution from the blended
H~{\sc i} $\lambda$3889 line are reliable, probably better than 20 per cent.
Although the strength of the He~{\sc i} $\lambda$3888 line could be much
affected by the effect of self-absorption due to a significant population on
the 2s\,$^{3}$S metastable level (e.g. \citealt{rob68}; \citealt{bss02}), the
majority of the line flux uncertainty is contributed by measurement errors.

Although the [O~{\sc iii}] $\lambda$4363 lines detected in the blue spectra
of PN1 and PN3 are faint (the S/N ratios $\sim$4--5), the O$^{2+}$/H$^{+}$
abundance derived from this line agrees with those derived from the [O~{\sc
iii}] $\lambda\lambda$4959 and 5007 nebular lines within errors. That was
expected because the abundances were derived assuming the electron
temperature yielded by the [O~{\sc iii}] line ratio. The
three [O~{\sc iii}] line detected in the spectra of both PN1 and PN3 yield
consistent O$^{2+}$/H$^{+}$ ionic abundances. The electron temperature of PN1
derived from the [O~{\sc iii}] $\lambda$4363 line is slightly higher than
that of PN3. It has been known that the weakly detected lines tend to be
over-measured when accurate sky subtraction is difficult \citep{kwi12}.
Considering the fact that the radial velocity of PN1 is lower than that of
PN3, and consequently accurate subtraction of the $\lambda$4358 mercury line
for PN3 is expected to be more difficult than for PN1, the relatively higher
electron temperature derived for PN1 indicates that the intensity of the
[O~{\sc iii}] $\lambda$4363 line of PN1 might be overestimated. Thus
we adopted an abundance ratio of 2.8$\times$10$^{-4}$, which was calculated
from the total intensity of the [O~{\sc iii}] $\lambda$4959 and 5007 lines,
as the O$^{2+}$/H$^{+}$ abundance for PN1. For PN3, that abundance ratio is
4.5$\times$10$^{-4}$. The total intensity of the [O~{\sc iii}]
$\lambda\lambda$4959 and 5007 lines in PN2 yields an abundance ratio of
1.8$\times$10$^{-4}$.

In both PN2 and PN3, the O$^{+}$/H$^{+}$ ionic abundances derived
from the $\lambda\lambda$3726 and 3729 lines agree with each other, which
was expected because the ionic abundances were determined based on the
electron density yielded by the [O~{\sc ii}] $\lambda$3726/$\lambda$3729 line
ratio. For PN1, the ionic abundances derived from the [O~{\sc ii}] lines
agree within errors, although slight difference is present. That is mainly
due to relatively large measurement uncertainty in the $\lambda$3729 line.
The O$^{+}$/H$^{+}$ abundance ratio derived from the $\lambda$3726 line was
adopted for all PNe. The N$^{+}$/H$^{+}$ abundances derived from the [N~{\sc
ii}] $\lambda\lambda$6548 and 6583 nebular lines detected in the red spectrum
of PN1 differ by about 40 per cent, while in PN2 and PN3 they differ by 19
and 10 per cent, respectively. Such significant difference in of PN1 is due
to a relatively large uncertainty in the faint [N~{\sc ii}] $\lambda$6548
line. Uncertainties in the N$^{+}$/H$^{+}$ ionic abundances of PN3 is
smaller because its red spectrum has better quality than the other two PNe.
We adopted the abundances calculated from the stronger $\lambda$6583 line as
the N$^{+}$/H$^{+}$ ionic abundances for the three PNe.

The [Ne~{\sc iii}] $\lambda$3967 line is blended with H~{\sc i} H7
$\lambda$3970, whose flux contribution was estimated from the observed
$\lambda$4340 line and the theoretical H~{\sc i} $\lambda$3970/$\lambda$4340
ratio. For all the three PNe, the corrected flux of the [Ne~{\sc iii}]
$\lambda$3967 line yields an Ne$^{2+}$/H$^{+}$ ionic abundance which is
generally consistent with the [Ne~{\sc iii}] $\lambda$3868 line. Given that
the $\lambda$3868 is stronger, the Ne$^{2+}$/H$^{+}$ abundance derived from
this [Ne~{\sc iii}] line is adopted for all three objects.
The S$^{+}$/H$^{+}$ and Ar$^{2+}$/H$^{+}$ abundance ratios were derived from
the [S~{\sc ii}] $\lambda$6731 and [Ar~{\sc iii}] $\lambda$7136 lines observed
in the red spectrum of PN3, respectively (Table\,\ref{ionic}).
%% Notes: At the physical condition of PN3 (Te = 10080 K and Ne = 1650 cm^-3),
%%%%%%%%  calculation of the five-level atomic model using equib shows that
%%%%%%%%  the [S II] 6716/6731 line ratio is about 0.76. Thus the [S II] 6716
%%%%%%%%  should be seen in PN3. The reason that it is not seen is probaly due
%%%%%%%%  to glitches in data reduction.

The uncertainties of abundances in the brackets following the abundance ratios
in Table\,\ref{ionic} were estimated based on two sources: (1) The measurement
uncertainties of line fluxes, and (2) the uncertainties in electron
temperatures resulted from the measurement uncertainties of the [O~{\sc iii}]
$\lambda$4363 auroral line. For PN1 and PN3, where the [O~{\sc iii}]
$\lambda$4363 line was detected in the spectra, uncertainties in the ionic
abundances of heavy elements are mainly contributed by the second source.
That is because emissivities of the heavy element CELs are very sensitive to
electron temperature under nebular conditions (e.g. \citealt{of06};
\citealt{liu12}), and thus the uncertainties of the derived ionic abundances
are, to a large extent, subjective to the uncertainties in CEL fluxes.
Although the electron density of PN1 was assumed to be 2000~cm$^{-3}$
(Table\,\ref{te_ne}), that does not affect the resultant ionic abundances
much, as discussed in Section\,\ref{part3:b}. For PN2, where the [O~{\sc iii}]
$\lambda$4363 line was not observed, a typical nebular electron temperature
of 10\,000~K was assumed. That may introduce significant uncertainties to the
ionic abundances of heavy elements, e.g. $\sim$20--30 per cent for some ions.
Measurement errors of line fluxes given in Table\,\ref{lines} are mainly
estimated from Gaussian-profile fits, i.e. difference between the directly
integrated flux and the flux given by Gaussian fits. Uncertainties of the
He$^{+}$/H$^{+}$ ionic abundances are mainly contributed by the measurement
errors in line fluxes because the emissivities of the He~{\sc i} recombination
lines are much less sensitive to the electron temperature compared to CELs.
The He$^{+}$/H$^{+}$ abundances derived from the He~{\sc i} $\lambda$3888 line
are of lower uncertainties than the other He~{\sc i} lines whose measurement
uncertainties are larger, although it is blended with the H~{\sc i}
$\lambda$3889 line.

The O$^{3+}$ ion needs to be taken into account when calculating the
O/H elemental abundances of our PNe. The determination of O/H requires the
He$^{2+}$/H$^{+}$ ionic abundance (because the ionization potential of He$^{+}$,
54.416~eV, is very close to that of the O$^{2+}$ ion, 54.934~eV) as derived
from the He~{\sc ii} $\lambda$4686 line, which is located at the end of the
blue CCD and thus difficult to be accurately measured due to very low S/N.
The upper limits of the $\lambda$4686 line intensities were estimated, which
yields ionization correction factors (ICFs) of oxygen for the three PNe close
to unity. %less than 1.15.
The ICF method was developed by \citet{kb94}.  Since only one
ionization stage of nitrogen ([N~{\sc ii}]) and neon ([Ne~{\sc iii}]) were
observed in our PN sample, the N/H elemental abundances were derived from the
O/H and O$^{+}$/H$^{+}$ abundances using Equations\,A1 and A2 in \citet{kb94},
and Ne/H was derived from O/H and O$^{2+}$/H$^{+}$ using Equations\,A28 and
A29 in \citet{kb94}. Only the S$^{+}$ ion was observed in PN3, and the total
abundance of sulphur was particularly uncertain. We used Equations\,A36, A37
and A38 in \citet{kb94} to estimate the S/H ratio for PN3. Only the Ar$^{2+}$
ion was observed in PN3, and thus the Ar/H elemental abundance was estimated
using Equations\,A32 and A33 in \citet{kb94}. Elemental abundances of N, O,
Ne, S and Ar are presented in Table\,\ref{element}, with uncertainties given
in brackets. Abundance uncertainties were derived directly from the
uncertainties in ionic abundances (Table\,\ref{ionic}), which were estimated
based on the measurement uncertainties of the line fluxes and electron
temperatures, as discussed earlier in this section. The uncertainties
introduced by ionization correction were also taken into account. Regardless
of the uncertainties in the ionization correction method, for all three PNe,
the oxygen abundances are the best calculated of all the heavy elements,
while uncertainties in the nitrogen and neon abundances are relatively larger
because they were derived from the ionic and total abundances of oxygen. The
sulphur and argon abundances are the most uncertain.

%%%% Table 4. Ionic abundances of the three NS PNe:
\begin{table*}
\caption{Ionic abundances.}
\label{ionic}
\centering
\begin{tabular}{llccc}
\hline
Ion & Line    & \multicolumn{3}{c}{Abundance (X$^{i+}$/H$^{+}$)}\\
\cline{3-5}\\
    & ({\AA}) &   PN1   &   PN2   &   PN3\\
\hline
He$^{+}$  & $\lambda$3888     & 0.106($\pm$0.027)  & 0.090($\pm$0.020)      & 0.097($\pm$0.015)     \\
          & $\lambda$4471     & 0.149($\pm$0.074)  &                        & 0.146($\pm$0.044)     \\
          & $\lambda$5876     &                    & 0.133($\pm$0.039)      & 0.095($\pm$0.023)     \\
          & $\lambda$6678     &                    &                        & 0.129($\pm$0.058)     \\
Adopted$^a$ &                 & 0.106($\pm$0.027)  & 0.090($\pm$0.020)      & 0.097($\pm$0.015)     \\
\\
N$^{+}$   & $\lambda$6548 & 3.9($\pm$1.1)$\times$10$^{-6}$ & 8.4($\pm$2.1)$\times$10$^{-6}$ & 1.2($\pm$0.5)$\times$10$^{-5}$\\
          & $\lambda$6583 & 5.4($\pm$1.0)$\times$10$^{-6}$ & 7.1($\pm$0.9)$\times$10$^{-6}$ & 1.1($\pm$0.1)$\times$10$^{-5}$\\
Adopted$^b$ &             & 5.4($\pm$1.0)$\times$10$^{-6}$ & 7.1($\pm$0.9)$\times$10$^{-6}$ & 1.1($\pm$0.1)$\times$10$^{-5}$\\
\\
O$^{+}$   & $\lambda$3726 & 3.3($\pm$0.7)$\times$10$^{-5}$ & 2.3($\pm$0.4)$\times$10$^{-5}$ & 3.3($\pm$0.6)$\times$10$^{-5}$\\
          & $\lambda$3729 & 2.9($\pm$1.1)$\times$10$^{-5}$ & 2.3($\pm$0.5)$\times$10$^{-5}$ & 3.4($\pm$0.7)$\times$10$^{-5}$\\
Adopted$^c$ &             & 3.3($\pm$0.4)$\times$10$^{-5}$ & 2.3($\pm$0.4)$\times$10$^{-5}$ & 3.3($\pm$0.6)$\times$10$^{-5}$\\
\\
O$^{2+}$  & $\lambda$4363 & 2.6($\pm$0.7)$\times$10$^{-4}$ &                                & 4.4($\pm$0.9)$\times$10$^{-4}$\\
          & $\lambda$4959 & 2.9($\pm$0.2)$\times$10$^{-4}$ & 1.8($\pm$0.1)$\times$10$^{-4}$ & 4.8($\pm$0.1)$\times$10$^{-4}$\\
          & $\lambda$5007 & 2.8($\pm$0.2)$\times$10$^{-4}$ & 1.8($\pm$0.1)$\times$10$^{-4}$ & 4.3($\pm$0.1)$\times$10$^{-4}$\\
Adopted$^d$ &             & 2.8($\pm$0.2)$\times$10$^{-4}$ & 1.8($\pm$0.1)$\times$10$^{-4}$ & 4.5($\pm$0.1)$\times$10$^{-4}$\\
\\
Ne$^{2+}$ & $\lambda$3868 & 5.8($\pm$0.5)$\times$10$^{-5}$ & 3.5($\pm$0.4)$\times$10$^{-5}$ & 1.3($\pm$0.1)$\times$10$^{-4}$\\
          & $\lambda$3967 & 5.6($\pm$0.9)$\times$10$^{-5}$ & 3.5($\pm$0.5)$\times$10$^{-5}$ & 1.3($\pm$0.1)$\times$10$^{-4}$\\
 Adopted$^e$ &            & 5.8($\pm$0.6)$\times$10$^{-5}$ & 3.5($\pm$0.4)$\times$10$^{-5}$ & 1.3($\pm$0.1)$\times$10$^{-4}$\\
\\
S$^{+}$   & $\lambda$6731 &                                &                                & 3.1($\pm$1.0)$\times$10$^{-7}$\\
\\
Ar$^{2+}$ & $\lambda$7136 &                                &                                & 1.1($\pm$0.3)$\times$10$^{-6}$\\
\hline
\end{tabular}
\begin{description}
\item [$^a$] The He$^{+}$/H$^{+}$ abundance ratio derived from the
$\lambda$3888 line is adopted.
\item [$^b$] The N$^{+}$/H$^{+}$ abundance derived from the [N~{\sc ii}]
$\lambda$6583 line is adopted.
\item [$^c$] For PN1, the O$^{+}$/H$^{+}$ abundance derived from the
$\lambda$3726 line is adopted.
\item [$^d$] The adopted O$^{2+}$/H$^{+}$ abundance is derived from the
total flux of [O~{\sc iii}] $\lambda\lambda$4959, 5007.
\item [$^e$] The Ne$^{2+}$/H$^{+}$ abundance derived from the [Ne~{\sc iii}]
$\lambda$3868 line is adopted.
\end{description}
\end{table*}

%%%% Table 5. Elemental abundances of the three NS PNe:
\begin{table*}
\caption{Elemental abundances$^a$.}
\label{element}
\centering
\begin{tabular}{lcccccc}
\hline
Element & \multicolumn{6}{c}{X/H}\\
\cline{2-7}\\
        & \multicolumn{2}{c}{PN1} & \multicolumn{2}{c}{PN2} & \multicolumn{2}{c}{PN3}\\
\hline
N       & 8.7($\pm$1.8)$\times$10$^{-5}$ & 7.94 & 1.0($\pm$0.4)$\times$10$^{-4}$ & 8.01 & 1.7($\pm$0.3)$\times$10$^{-4}$ & 8.23\\
O       & 3.6($\pm$0.5)$\times$10$^{-4}$ & 8.55 & 2.6($\pm$0.5)$\times$10$^{-4}$ & 8.42 & 5.2($\pm$0.6)$\times$10$^{-4}$ & 8.72\\
Ne      & 7.4($\pm$1.4)$\times$10$^{-5}$ & 7.87 & 5.8($\pm$1.4)$\times$10$^{-5}$ & 7.76 & 1.5($\pm$0.3)$\times$10$^{-4}$ & 8.16\\
S       &                                &      &                                &      & 4.8($\pm$2.1)$\times$10$^{-6}$ & 6.68\\
Ar      &                                &      &                                &      & 2.0($\pm$0.8)$\times$10$^{-6}$ & 6.31\\
\hline
\end{tabular}
\begin{description}
\item [$^a$] The second column of abundances for each PN are in logarithm,
12\,+\,$\log_{\rm 10}${(X/H)}.
\end{description}
\end{table*}

\section{\label{part4}
Discussion}

\begin{figure}
\begin{center}
\includegraphics[width=8.5cm,angle=-90]{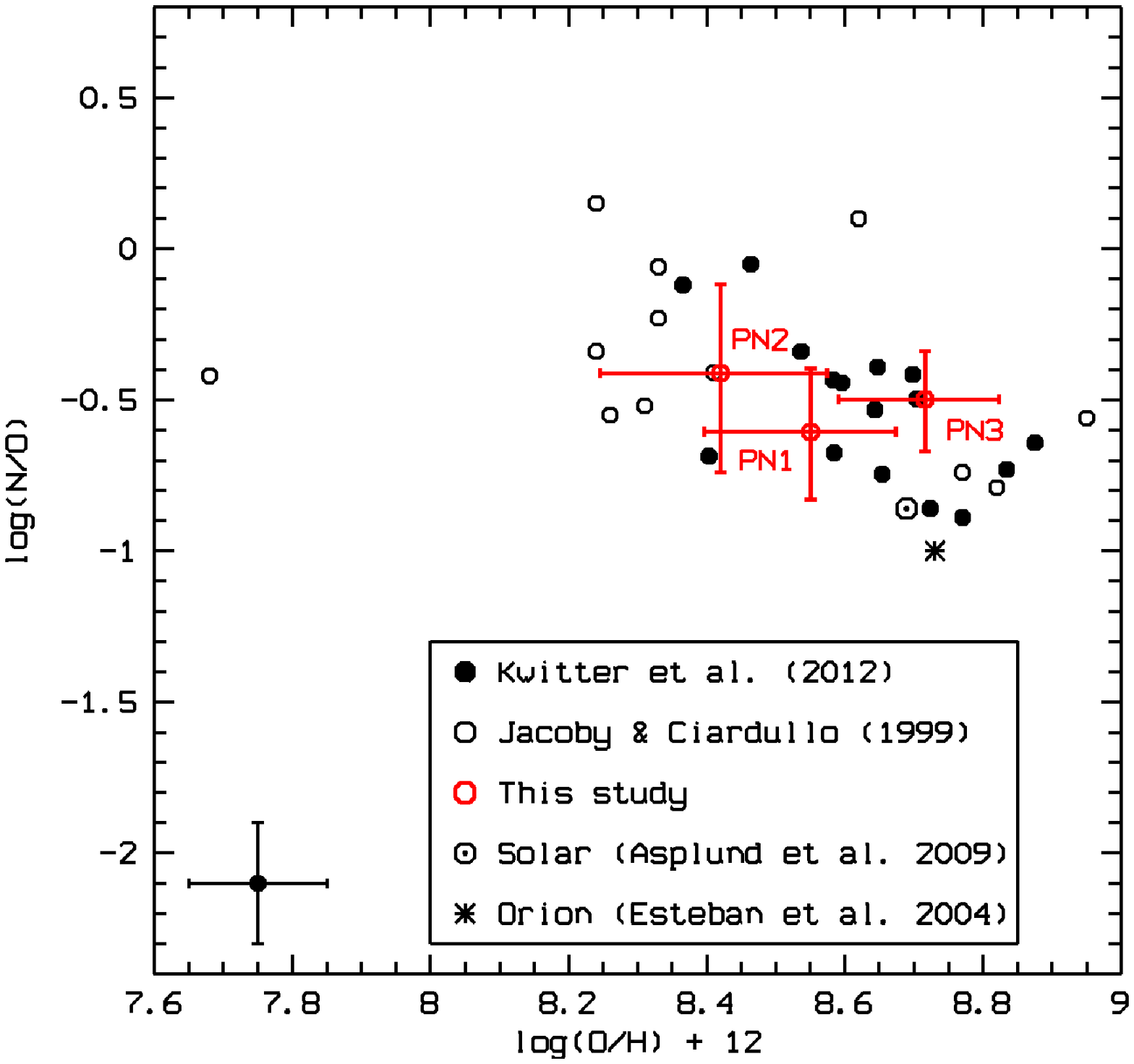}
\caption{The N/O versus the O/H abundance ratio in the logarithmic scale.
Different symbols represent different data sources (see the legend in the
diagram): The red circles are the three Northern Spur PNe in this
study, black filled circles are the M31 disk PNe from \citet{kwi12}, and
black open circles are the M31 bulge and disk PNe from \citet{jc99}; the
solar value is from \citet{asp09}, and the asterisk is the Orion nebular
abundance from \citet{est04}. Error bars are given for the three Northern Spur
PNe. Representative error bars of the M31 disk PNe observed by \citet{kwi12}
are given in the lower-left corner.}
\label{NOratio_O}
\end{center}
\end{figure}

\begin{figure}
\begin{center}
\includegraphics[width=8.5cm,angle=-90]{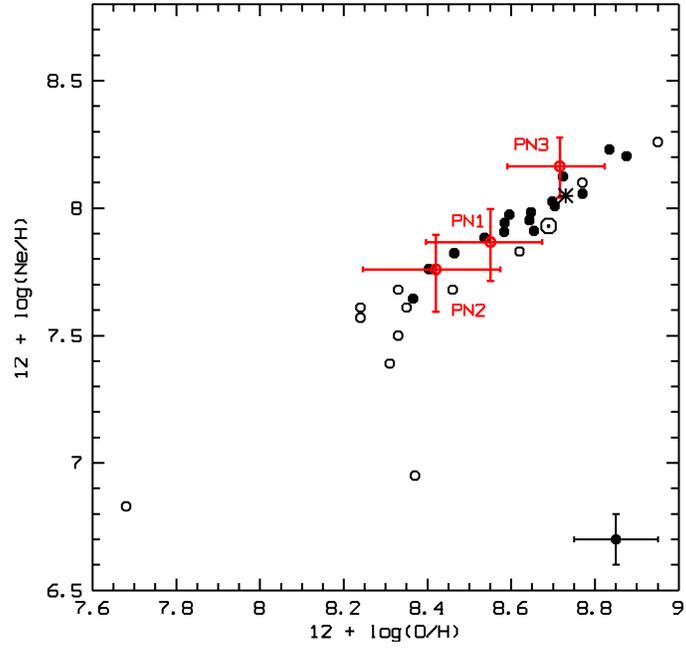}
\caption{Same as Figure\,\ref{NOratio_O} but for Ne/H versus O/H.
Representative error bars of the M31 disk PNe observed by \citet{kwi12} are
given in the lower-right corner.}
\label{Ne_O}
\end{center}
\end{figure}

\begin{figure}
\begin{center}
\includegraphics[width=8.5cm,angle=-90]{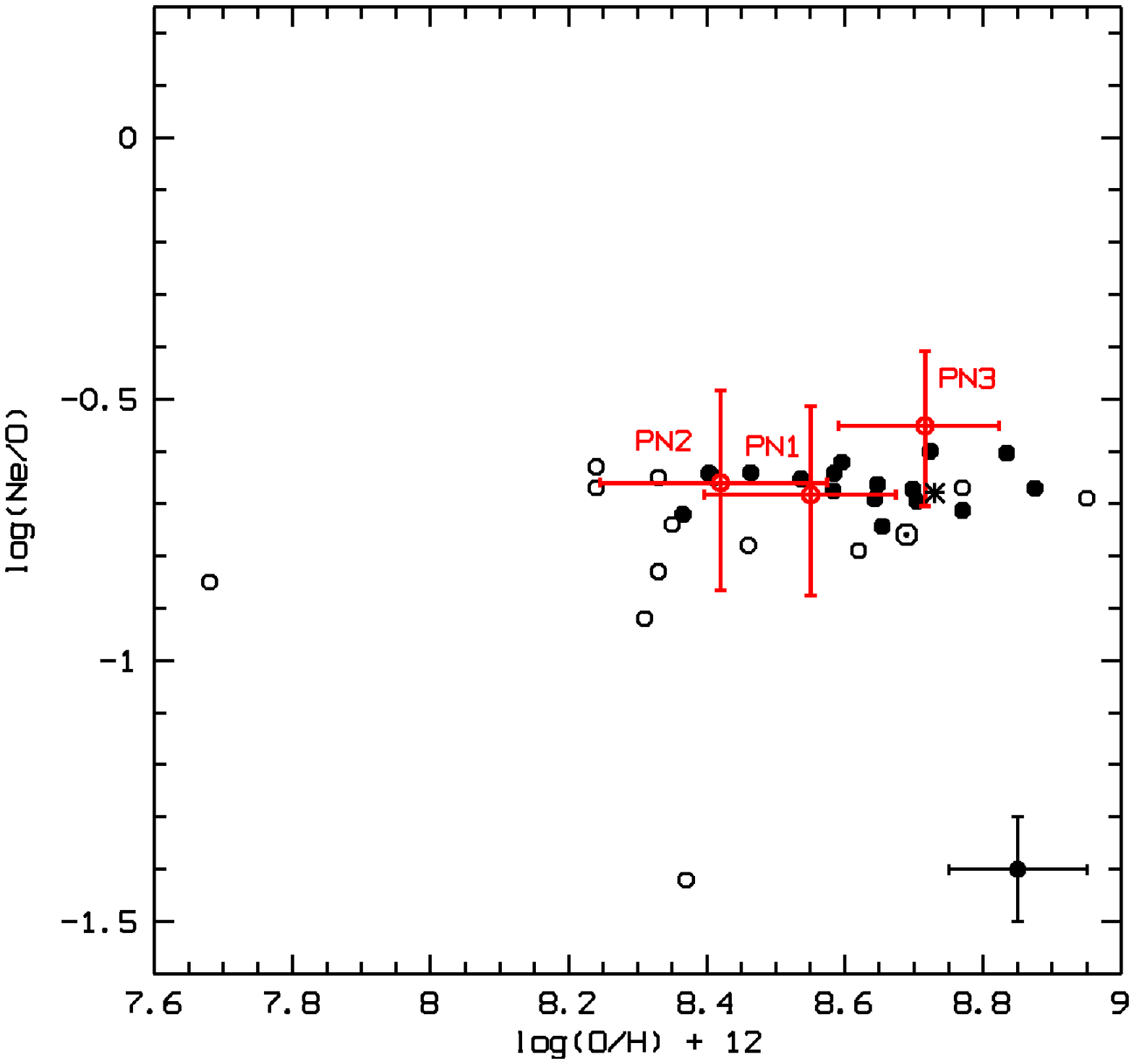}
\caption{Same as Figure\,\ref{Ne_O} but for Ne/O versus O/H.}
\label{NeO_O}
\end{center}
\end{figure}

\begin{figure}
\begin{center}
\includegraphics[width=8.5cm,angle=-90]{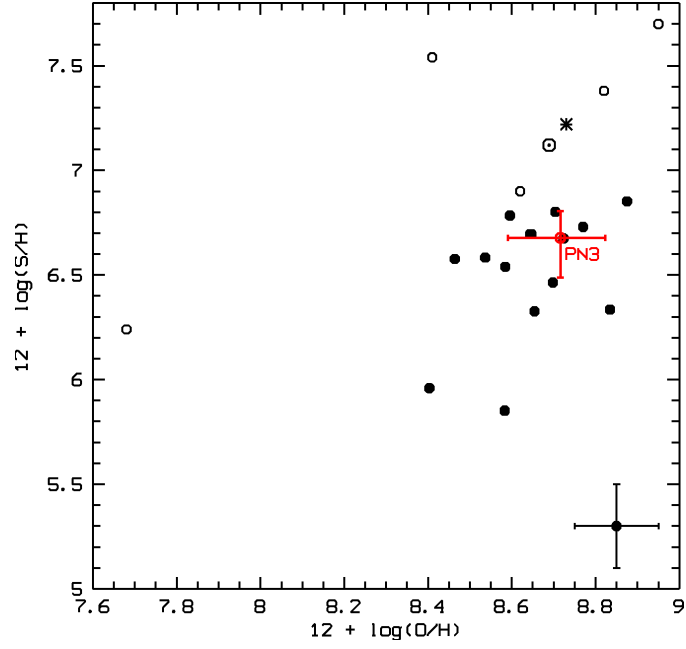}
\caption{Same as Figure\,\ref{Ne_O} but for S/H versus O/H. The S$^{+}$ ion
is only detected in the spectrum of PN3.}
\label{S_O}
\end{center}
\end{figure}

\begin{figure}
\begin{center}
\includegraphics[width=8.5cm,angle=-90]{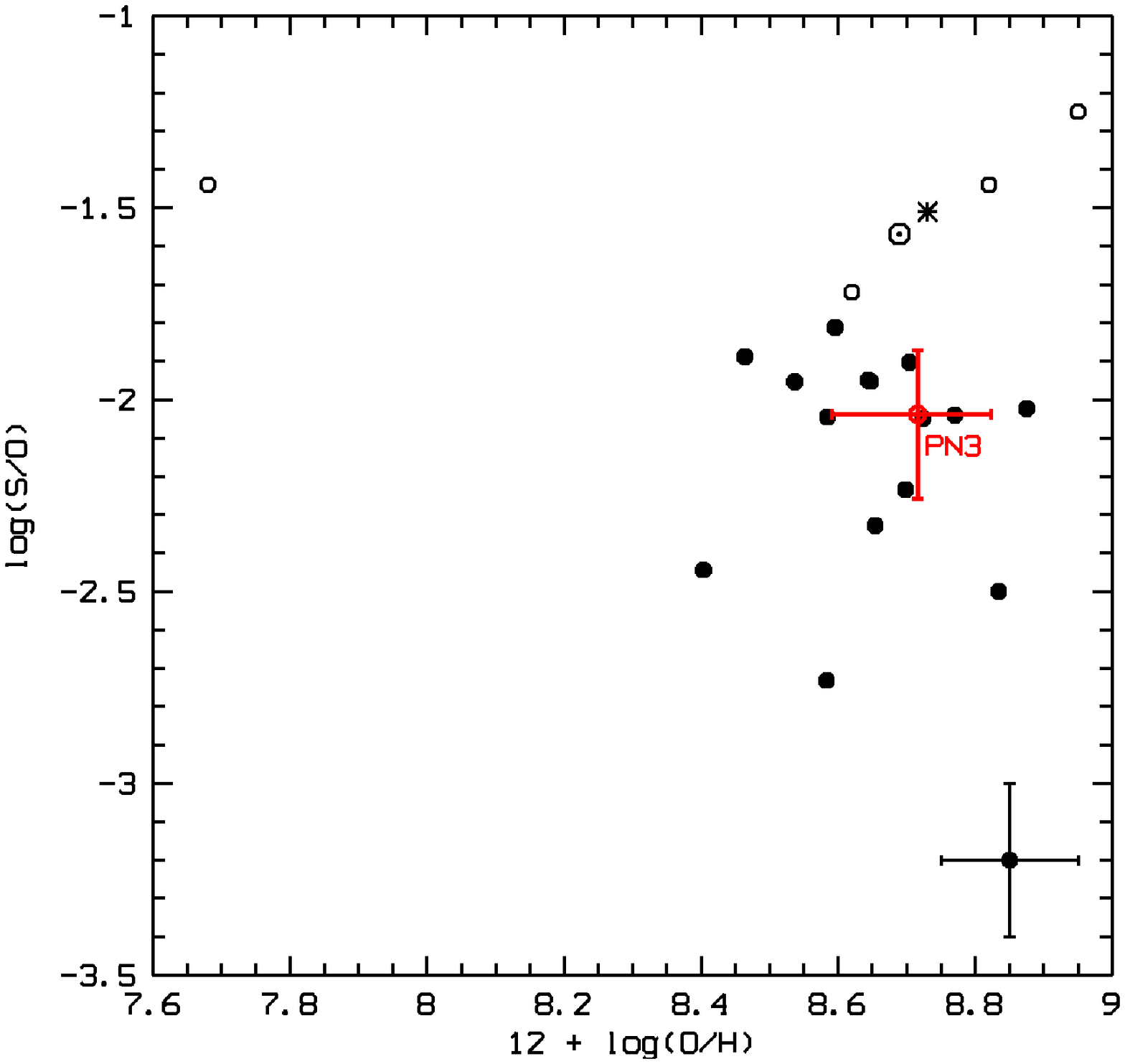}
\caption{Same as Figure\,\ref{S_O} but for S/O versus O/H.}
\label{SO_O}
\end{center}
\end{figure}

\begin{figure}
\begin{center}
\includegraphics[width=8.5cm,angle=-90]{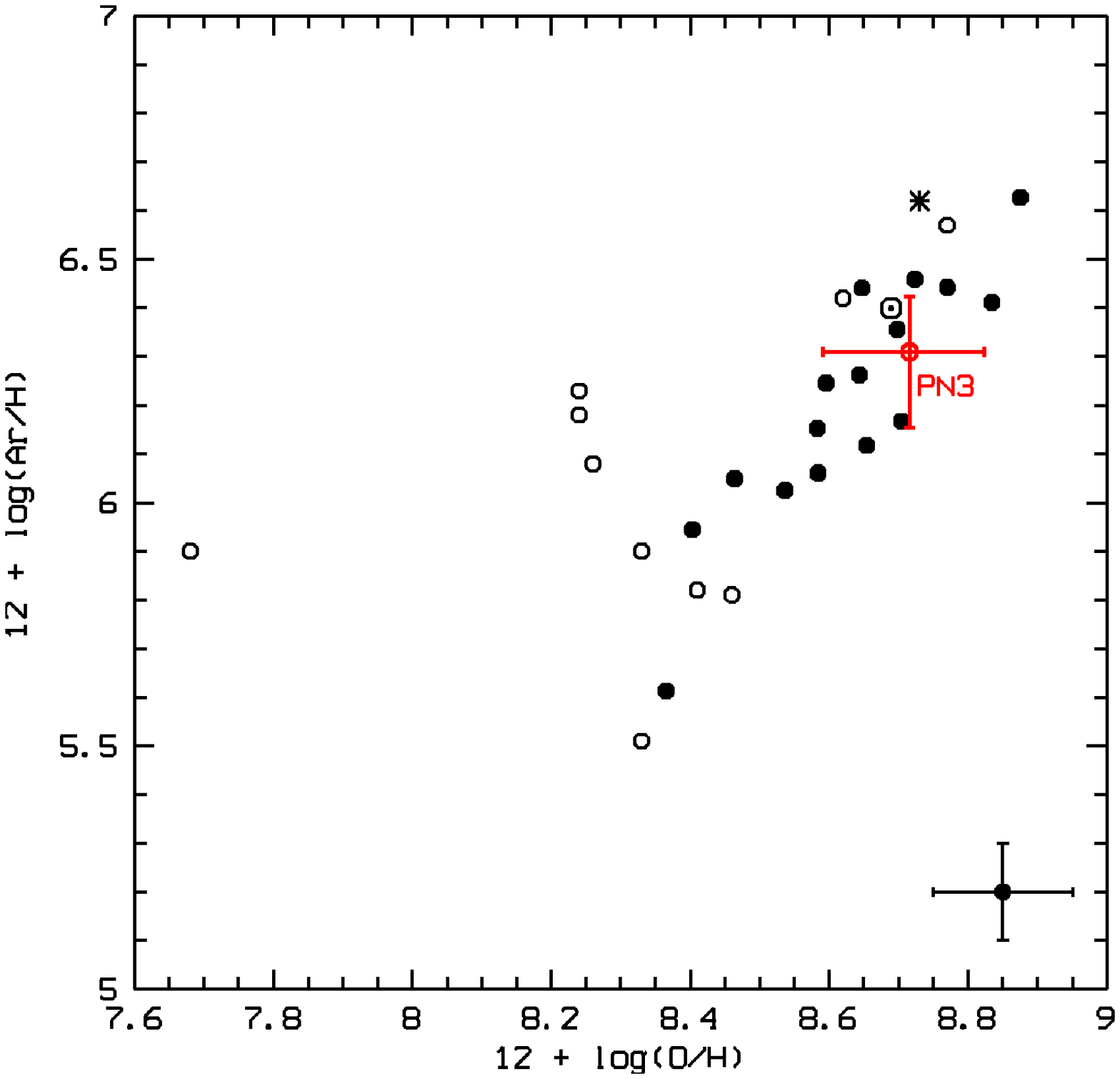}
\caption{Same as Figure\,\ref{S_O} but for Ar/H versus O/H.}
\label{Ar_O}
\end{center}
\end{figure}

\begin{figure}
\begin{center}
\includegraphics[width=8.5cm,angle=-90]{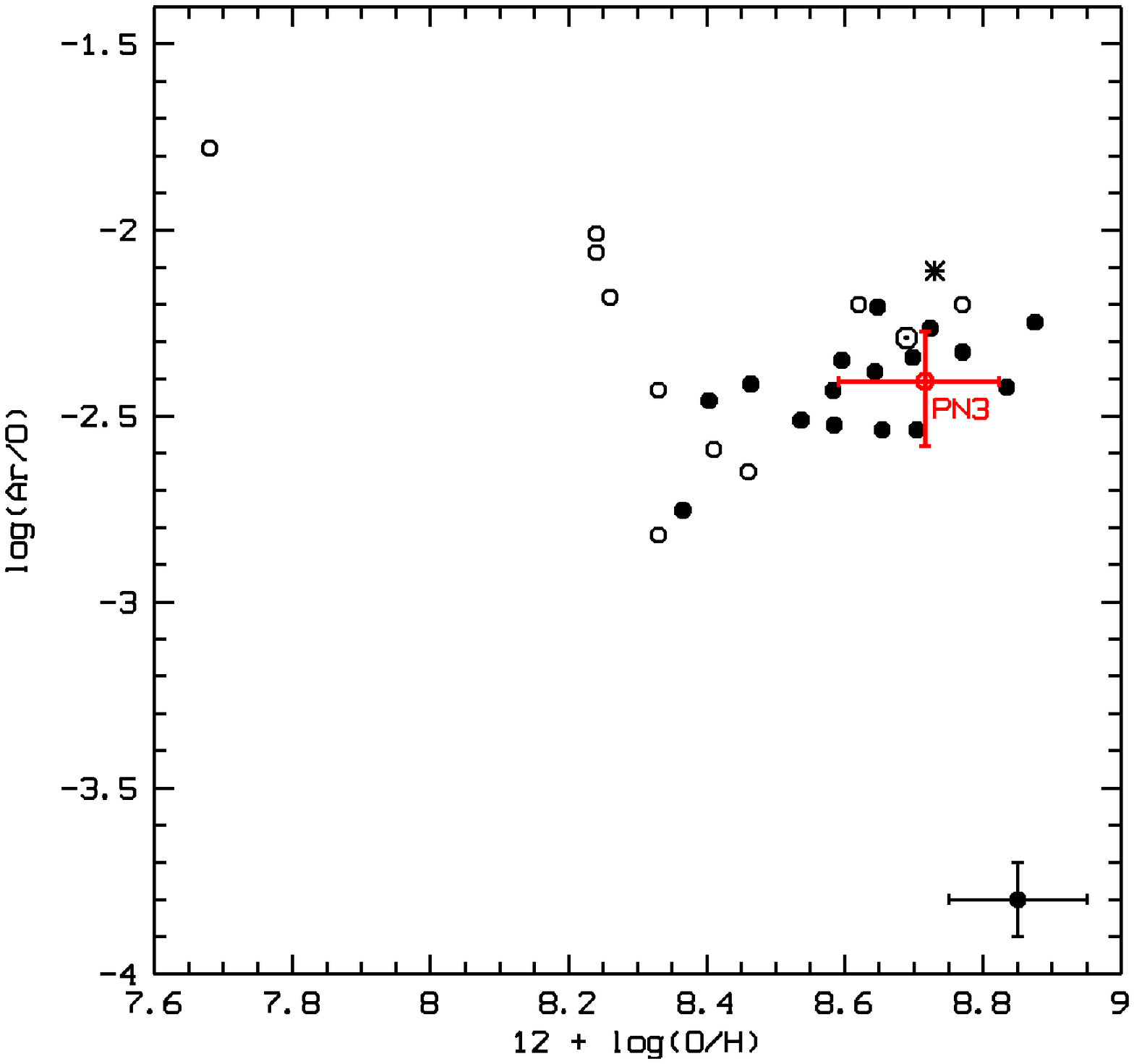}
\caption{Same as Figure\,\ref{S_O} but for Ar/O versus O/H.}
\label{ArO_O}
\end{center}
\end{figure}

Gaseous nebulae, mainly PNe and H~{\sc ii} regions, are useful probes of the
past chemical composition of the interstellar medium (ISM). The $\alpha$-element
abundances of a PN reflect those in the ISM at the time when the progenitor
star formed, while the $\alpha$-element abundances of a sample of H~{\sc ii}
regions provide a `snapshot' of the current status of chemical evolution of
galaxies. Comparison of the abundances of PNe and H~{\sc ii} regions on the
disk of a spiral galaxy helps to study the chemical history and production
processes of the elements. Furthermore, studying the relations between
abundances of different $\alpha$-elements helps to constrain the production
processes and the relative yields of each element. Ratios of different
$\alpha$-element abundances also reflect the enrichment by the progenitor
stars.  Figures\,\ref{NOratio_O}--\ref{ArO_O} show the abundance
correlations in our sample as well as the M31 disk and bulge PNe from the
literature, in the logarithmic scale.  Figure\,\ref{NOratio_O} presents the
$\log$(N/O) versus $\log$(O/H) abundance relation. Also overplotted are the
M31 disk PNe from \citet{kwi12} and the M31 bulge and disk sample observed by
\citet{jc99}. Also presented in the plot are the solar values from
\citet{asp09} and the nebular abundances of Orion from \citet{est04}.
Figure\,\ref{NOratio_O} shows that there is no obvious trend in the
$\log$(N/O) versus $\log$(O/H) ratio in our sample.
%Comparison with previous observations of the $\log$(N/O) ratios of PNe in
%both M31 and the MW (see Figure\,8 of \citealt{kwi12}) indicates that our
%three PNe are more close to the Type~II classification, similar to the M31
%disk PNe observed by \citet{kwi12}. \citet{kb94} defines Type~I PNe as
%those objects that have N/O ratio higher than 0.8, using spectrophotometric
%observations of MW PNe. The progenitor stars of Type~I PNe have experienced
%envelope-burning conversion to nitrogen of dredged-up carbon. Such nebulae
%are recognized by having nitrogen abundances that exceed the total C\,+\,N
%abundance of H~{\sc ii} regions in the same galaxy. Type~II PNe have
%relatively less massive progenitor stars and N/O ratios are lower. The N/O
%ratios of our Northern Spur sample are 0.25, 0.39 and 0.33 (PN1, PN2 and
%PN3, respectively). Thus according to the \citet{kb94}, they are definitely
%non-Type~I.  If that is true, then given the low-mass progenitors of our
%sample, they could be old populations that might originate from one of the
%dwarf satellites of M31 during evolution in the distant past. Origin of the
%Northern Spur is discussed later in this section.

Figure\,\ref{Ne_O} displays a positive correlation between $\log$(Ne/H) and
$\log$(O/H). That confirms the tight relation between these two elements,
which have by far been observed for PNe in both the MW and M31. The neon-oxygen
abundance distribution of our sample agree with the slope of other M31 PNe
within the errors. Figure\,\ref{NeO_O} shows that there is no obvious
correlation between the Ne/O ratio and O/H. Figure\,\ref{S_O} shows
$\log$(S/H) versus $\log$(O/H), and Figure\,\ref{SO_O} is $\log$(S/O) versus
$\log$(O/H). The sulphur-oxygen correlation shows larger scatter than neon
(Figure\,\ref{Ne_O}). As pointed out by \citet{kwi12} as well as by earlier
studies, the determination of sulphur abundances in the PNe of M31 is very
challenging: 1) Ionization correction introduces uncertainties because of the
ions that can not be observed in the optical, e.g. S$^{3+}$; and 2) the lines
from both S$^{+}$ and the predominant ionization state S$^{2+}$ are faint, or
absent, in the M31 PNe. Our observations (in PN3) confirm the results of
\citet{kwi12} that derived sulphur abundances for the M31 PNe are lower than
that of the Sun. The [S~{\sc ii}] line is only observed in PN3, and
Figures\,\ref{S_O} and \ref{SO_O} show that our abundance ratios lie within
the ranges of \citet{kwi12}. The behavior of argon is similar to that of
neon, as shown in Figure\,\ref{Ar_O}, and PN3 is located well within the
argon-oxygen correlation of \citet{kwi12}. There is no obvious correlation
between $\log$(Ar/O) and $\log$(O/H), as indicated by Figure\,\ref{ArO_O}.

\begin{figure*}
\begin{center}
\includegraphics[width=12cm,angle=-90]{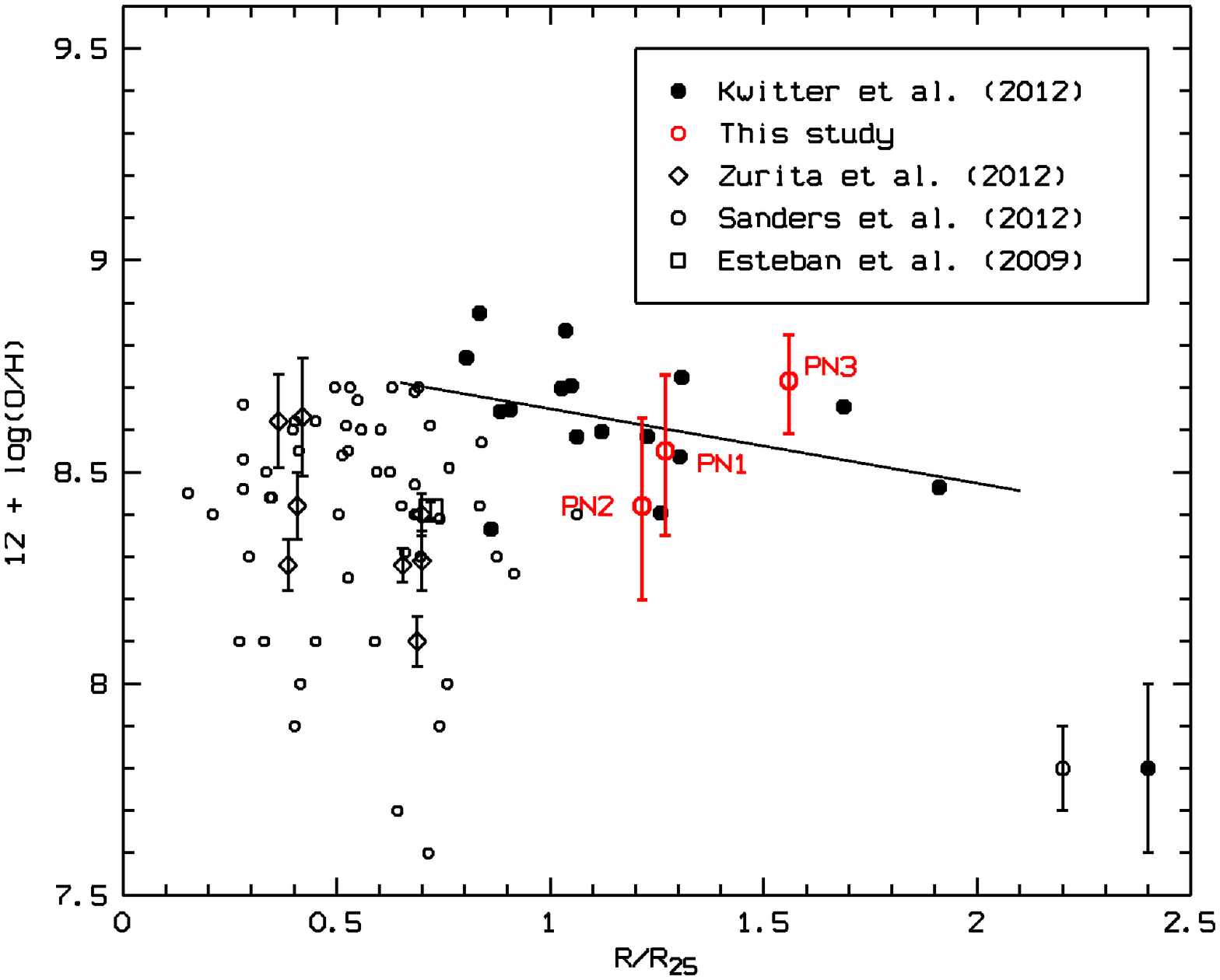}
\caption{Oxygen gradient of the PNe and H~{\sc ii} regions in M31. The
$R_{\rm 25}$ value of M31 is 22.4~kpc as adopted from \citet{ggh98}. The black
filled circles are the M31 disk PNe observed by \citet{kwi12}, the black open
circles are the M31 PNe from \citet{san12}. The open diamonds are nine H~{\sc
ii} regions on the disk of M31 observed by \citet{zb12}, and the claimed
abundance uncertainties are given as error bars. The open square is an M31
H~{\sc ii} region (K932) observed by \citet{est09}. The three red open circles
are our current observations, and the abundance uncertainties are given. The
galactocentric distances of the three Northern Spur PNe in M31 have been
rectified from the sky-projected distances. Representative abundance
uncertainties of \citet{kwi12} and \citet{san12} are indicated in the
$lower$-$right$ corner. The black straight line is a simple linear regression
fit to the M31 disk PNe (the black filled circles) observed by \citet{kwi12}.}
\label{gradient}
\end{center}
\end{figure*}

The distribution of oxygen abundance with the galactocentric distance of the
M31 PNe and H~{\sc ii} regions is shown in Figure\,\ref{gradient}. The M31
disk sample observed by \citet{kwi12} and our three Northern Spur PNe are both
given in the plot. Also presented are the M31 PNe observed by \citet{san12}
and nine H~{\sc ii} regions on the disk of M31 observed by \citet{zb12}. All
galactocentric distances have been rectified for the effects of projection
on the sky plane and reduced in units of $R_{25}$ which is 22.4~kpc for M31
\citep{ggh98}. The galactocentric distances (in kpc) of our three PNe have
been rectified using the formula

\begin{equation}
\label{rectify}
R_{\rm rectified} = (X^{2} + (Y/{\rm cos}i)^{2})^{1/2}
\end{equation}
given by \citet{kwi12}. Here we assumed that the three Northern Spur PNe are
all located on the disk of M31. $X$ and $Y$ are distances to the galactic
center projected on the major and minor axes, respectively. The inclination
angle $i$ of the M31 disk to the plane of the sky is 77.7$^{\rm o}$ as adopted
from \citet{dev58}. The $X$ and $Y$ values were calculated from the RA and Dec
of a PN as well as the position angle (PA) of the M31 main axis (37.7$^{\rm
o}$; \citealt{dev58}).

\begin{figure*}
\begin{center}
\includegraphics[width=15cm,angle=0]{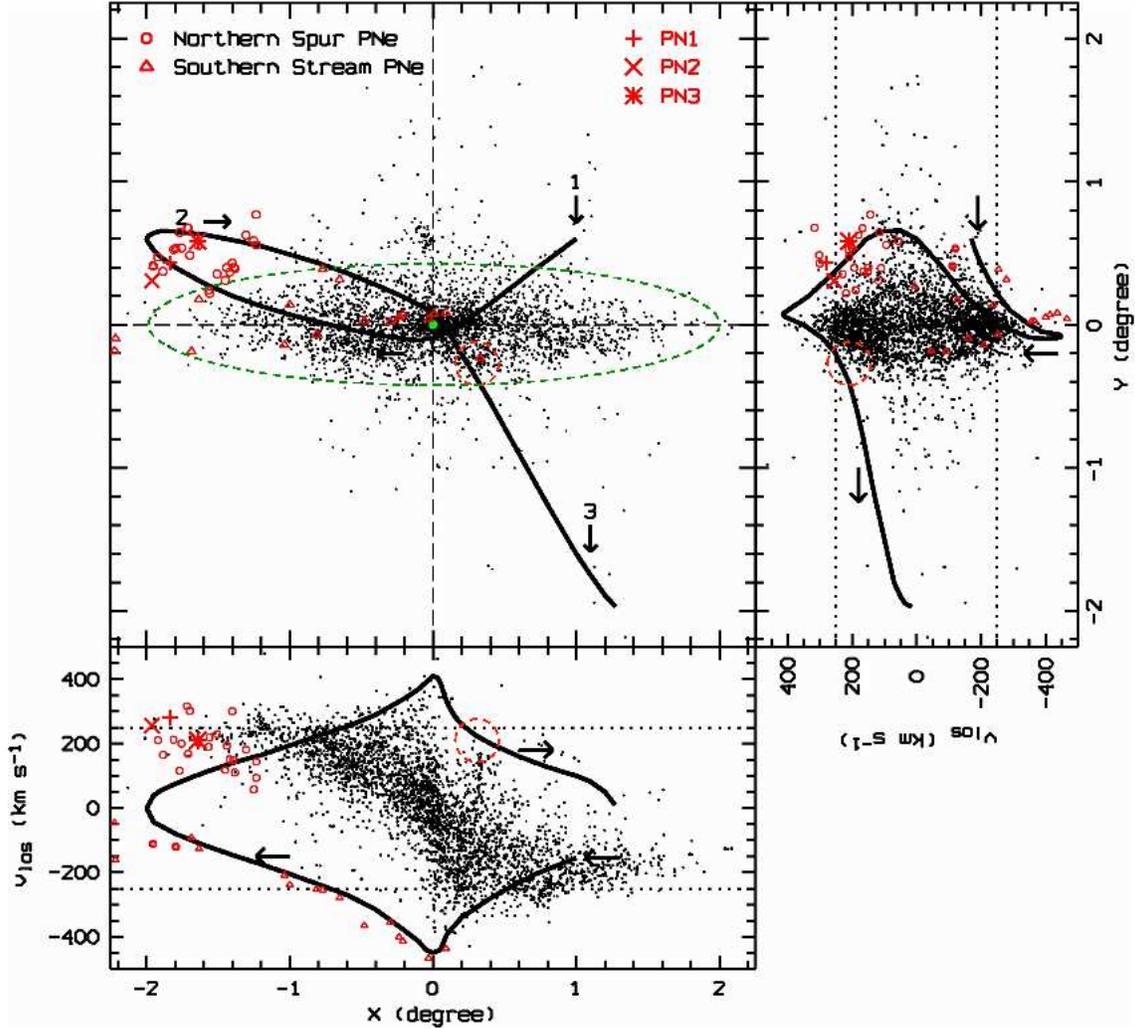}
\caption{The spatial distribution and kinematics of PNe in M31 (see
description of the $X$ and $Y$ coordinates in the text). Samples are from
\citet{mer06}. Small red circles represent the Northern Spur PNe and red
triangles are those identified by \citet{mer06} as forming a continuation
of the Southern Stream. The three Northern Spur PNe studied in the
current paper are highlighted. In the $upper$ main panel: Overplotted is the
orbit (thick black curve) proposed by \citet{mer03}; green dot represents the
center of M31, and the green dashed ellipse has a semimajor axis of 2$^{\rm
o}$ ($\sim$27~kpc) and represents a disk with an inclination angle $i$ =
77.7$^{\rm o}$ \citep{dev58}. In the two side panels ($lower$ and $right$):
Projection of the orbit in line-of-sight velocity with respect to M31,
$v_{\rm los}$, versus distance along the major and minor axes of M31 is
superimposed on the PNe data. Arrows show direction along the stream. The red
dashed circle shows the location of M32. Format of the figure follows
Figure\,$2$ in \citet{mer03}, and the orbit is reproduced based on that
figure (with permission of the authors).}
\label{orbit}
\end{center}
\end{figure*}

It can be seen from Figure\,\ref{gradient} that the oxygen abundances of the
M31 disk PNe observed by \citet{kwi12} are generrally higher than those of
\citet{san12}. However, the radial distribution of the sample of \citet{san12}
is more restricted. In our sample, PN3 has a higher oxygen abundance than the
M31 disk PNe at similar galactocentric distances. This PN also has the
best-quality spectrum among the three. That indicates our sample, at least
PN3, might be different from the M31 disk population.  Abundances of the
sample observed by \citet{san12} show very large scatter and systematically
lower than those of \citet{kwi12}. The nine H~{\sc ii} regions in the disk
of M31 observed by \citet{zb12} were derived from the direct
$T_\mathrm{e}$-based method. These H~{\sc ii} regions seem to have steeper
oxygen gradient than \citet{kwi12}, but they are spatially restricted and
too few sample. Judging from the spatial distribution of our three
Northern Spur PNe (Figure\,\ref{orbit}), PN3 may be more associated with this
substructure, although all three PNe have already been identified by
\citet{mer06} as belonging to the Northern Spur; the other two PNe are located
more close to the major axis of M31.  If the orbital model of \citet{mer03}
is correct, i.e. the Northern Spur of M31 is connected to the Southern Stream,
our observations of PN3 seem to be in line with the postulation that
the Northern Spur substructure is composed of the tidal debris of the M31's
satellite galaxies.
%However, more observations in the Northern Spur and in the Southern Stream
%are needed.

%% fangx: Discussion of the possible origins of Northern Spur here:
The origin of Northern Spur is still largely unclear, although large-area
photometric observations of M31 had been carried out more than 10 years ago.
Previous studies of the substructure in M31 have revealed that Northern Spur
is metal-rich. However, those results are only based on color information.
Quantitative spectroscopy are needed to confirm that nature. \citet{mer03} was
among the first to formally propose the possible origin of Northern Spur, i.e.
it might be associated with the Southern Stream, although that had been
inferred by \citet{fer02} and \citet{mccon03}. Using the kinematic information
of PNe in the disk of M31 and based on the studies of \citet{fer02} and
\citet{mccon03}, \citet{mer03} constructed an orbit model of the stellar
stream that connects the Northern Spur to the Southern Stream. In this model,
the area of the Northern Spur encompasses the turning point of the
orbit which is strongly warped near the center of M31 (see Figure\,$2$
of \citealt{mer03}; see also Figure\,\ref{orbit} in this paper which is
constructed based on Figure\,2 of \citealt{mer03}, with permission of the
authors). Judging from the projected position and kinematics of M32,
\citet{mer03} hypothesized that this satellite might be a parent of the
stream, although the exact position of M32 with respect to M31 is still an
open question.

Figure\,\ref{orbit} shows this orbit in the $X$-$Y$ coordinate system in an
M31-based reference frame, where $X$ lies along the major axis of M31 and
increases toward the south-west, and $Y$ lies along the minor axis and
increases toward the north-west. Both coordinates are calculated following
the geometric transformations of \citet{hbk91}. Also presented in
Figure\,\ref{orbit} is the projection of this orbit in the line-of-sight
velocity with respect to to M31, $v_{\rm los}$, versus distance along the
major and minor axes. Spatial and kinematic distribution of the M31
PNe observed by \citet{mer06} are presented along with the orbit.
PNe in the region of Northern Spur identified by \citet[see Figure\,32
therein]{mer06} as well as those identified by the same authors as forming a
continuation of the Southern Stream are highlighted with different symbols
in Figure\,\ref{orbit}. The three Northern Spur PNe studied in the current
paper are also highlighted.  In the lower panel of Figure\,\ref{orbit} (i.e.
$v_{\rm los}$ versus $X$), dispersion in the line-of-sight velocities of the
Northern Spur PNe with respect to the projected orbit is relatively large,
indicating the orbit in this section might be of large uncertainty. That is
expected because the orbit model of \citet{mer03} is based on a limited
number of PNe ($\sim$20) and relatively simple assumptions.
Considering the fact that PN3 in our sample has relatively higher
oxygen abundance than the M31 disk sample at similar galactocentric distances
(Figure\,\ref{gradient}), our observations seem to be in line with the
postulation of \citet{mer03}.  Also noticeable in Figure\,\ref{orbit} (also
Figure\,$32$ in \citealt{mer06}) is that the 20 PNe (two of them might be the
Northern Spur candidates, as indicated by \citealt{mer06}) associated with
the Southern Stream generally well fit the orbit model. Future spectroscopy
of these PNe will help to confirm their true nature.  The kinematics
of PNe in the Northern Spur region are indistinguishable from those of the
disk, and \citet{mer06} suggests that this substructure is related to the
disk, perhaps indicative of a warp. However, deep spectroscopy of more PNe
in this region is definitely needed so that they can be chemically
distinguished from those in the disk. Figure\,\ref{orbit} here is to visually
demonstrate the position of the Northern Spur PNe relative to the model orbit
of \citet{mer03}.
As to the origin of the Northern Spur, observations of more PNe
therein and the PNe in the Southern Stream may help to confirm whether these
two substructures linked. Observations of the PNe M32 may also help to
understand the interaction between M32 and M32 and assess whether the Northern
Spur and Southern Stream both originate from M32. Currently, observations of
PNe in M32 are scarce \citep{rsm99} and results inconclusive. More high-quality
spectroscopic observations are preferred.

%%%% Figure 9. N/O ratio versus oxygen:

%%%% Figure 10. Neon versus oxygen:

%%%% Figure 11. Ne/O ratio versus oxygen:

%%%% Figure 12. Sulfur versus oxygen:

%%%% Figure 13. S/O versus oxygen:

%%%% Figure 14. Argon versus oxygen:

%%%% Figure 15. Ar/O versus oxygen:

%%%% Figure 16. The oxygen gradient:

%%%% Figure 17. The orbit model of Merrett et al. (2003):

\section{\label{part5}
Summary and conclusion}

We present deep spectroscopy of three PNe in the Northern Spur of M31 using
DBSP on the 5.1\,m Hale Telescope at the Palomar Observatory. The
sample is selected from \citet{mer06}. This is the first chemical study of
PNe in this substructure. The [O~{\sc iii}] $\lambda$4363 auroral line is
detected in the spectra of two objects after meticulous work on the
subtraction of sky background. Electron temperatures were determined for two
PNe. Ionic abundances of heavy elements were derived from the [N~{\sc ii}],
[O~{\sc iii}], [Ne~{\sc iii}], [S~{\sc ii}] and [Ar~{\sc iii}] CELs detected
in the spectra. The N/H, O/H, Ne/H, S/H and Ar/H elemental abundances were
estimated. Correlations between oxygen and $alpha$-element abundance ratios
were studied, using our sample and other M31 PNe from the literature. One of
the three Northern Spur PNe has relatively higher oxygen abundance than both
the M31 disk PNe at similar galactocentric distances and the average abundance
of the M31 disk sample. Our study is thus seems to be in line with
the postulation that the Northern Spur might be connected to the Southern
Stream and are metal-enriched.  More observations of PNe at different
substructures, in combination with the kinematic information, are needed to
assess the properties of those substructures and further constrain the
possible origins of the Northern Spur.

\section*{Acknowledgements}
This research is based on data obtained using the 5.1\,m Hale Telescope at
Palomar Observatory owned and operated by the California Institute of
Technology. This project is supported by the Telescope Access Program (TAP),
which was initiated in 2011 and aims to provide the astronomers based in
China more access to leading facilities at a range of apertures. The project
is also supported by the National Science Foundation of China (No. 10933001).
XF and YZ thank the staff of Palomar Observatory for kind help during the
observations. YZ thanks the Research Grants Council of the Hong Kong Special
Administrative Region, China for financial support (Grants HKU7073/11P). RGB
acknowledges support from MICINN AYA2010-1508. We thank Martin A. Guerrero
and Enrique P\'{e}rez for valuable comments and suggestions. We also thank
Michael Merrifield for giving us the permission to produce a figure
(Figure\,\ref{orbit} in this paper) based on Figure\,$2$ in \citet{mer03}.
We also would like to thank an anonymous referee whose comments have greatly
improved the quality of this article.

\end{document}